\documentclass[12pt]{article}

\usepackage[usenames,dvipsnames]{xcolor}

\usepackage{tikz}
\usetikzlibrary{matrix,arrows,positioning,shapes,snakes,fadings,decorations.pathmorphing,
decorations.pathreplacing,automata,shadows,decorations.markings,patterns,decorations.text}
\usepackage{jheppub}
\usepackage{amsmath,amssymb,euscript,array,mathrsfs,appendix,ctable,mathabx}
\usepackage{mathtools,float}
\usepackage{arydshln}
\usepackage{todonotes}
\usepackage{graphicx}
\usepackage{wrapfig}
\usepackage{enumitem}
\usepackage{multirow}

\newcommand*\circled[1]{{\footnotesize\tikz[baseline=(char.base)]{%
            \node[shape=circle,fill=black!20,draw,inner sep=1pt] (char) {#1};}}} 
 
\newcommand\SBH{S_\text{BH}}
\newcommand\ZZ{\EuScript S}

\newcommand \arXiv [1]{\href{http://arxiv.org/abs/#1}{\tt arXiv:#1}} 

\usepackage{titlesec}
\titleformat{\subsection}[display]{\it}{}{0.1cm}{\vspace{-1.5cm}\begin{center}\thesubsection\hspace{0.2cm}}[\end{center}\vspace{-0.5cm}]

\newcommand{\ket}[1]{|#1\rangle}

\newcommand{\ext}{\text{ext}}

\newcommand{\EQ}[1]{\begin{equation}\begin{split} #1
\end{split}\end{equation}}

\title{Islands in the Stream of Hawking Radiation}
\author{Timothy J. Hollowood, S.~Prem Kumar, Andrea Legramandi and Neil Talwar}
\affiliation{Department of Physics, Swansea University, Swansea, SA2 8PP, U.K.}
\emailAdd{t.hollowood@swansea.ac.uk,s.p.kumar@swansea.ac.uk,\\ andrea.legramandi@swansea.ac.uk,n.talwar.2017429@Swansea.ac.uk}
\abstract{
We consider the island formula for the entropy of subsets of the Hawking radiation in the adiabatic limit where the black hole evaporation is very slow. We find a simple concrete `on-shell' formula for the generalized entropy which involves the image of the island out in the stream of radiation, the `island in the stream'. The resulting recipe for the entropy allows us to calculate the quantum information properties of the radiation and verify various constraints including the Araki-Lieb inequality and strong subadditivity. 
}

\notoc
\begin{document}

\maketitle

\newpage

\tableofcontents

\section{Introduction}

There has recently been a huge leap forward in understanding the information loss paradox of black holes  \cite{Penington:2019kki,Almheiri:2019qdq}. We can now pinpoint exactly the missing element in Hawking's original calculation \cite{Hawking:1974sw,Hawking:1976ra} of the state of the radiation emitted by the black hole. Somewhat unexpectedly, the missing ingredient is already contained within the  semi-classical framework of gravity.  Specifically, there are additional saddle points, the replica wormholes, of the functional integral  for  the QFT  used for computing the von Neumann entropy of the radiation in the semi-classical limit \cite{Penington:2019kki,Almheiri:2019qdq,Almheiri:2020cfm}. Hawking's calculation 
is still valid when the black hole is young and the entropy in the radiation increases monotonically, but when the black hole is old, beyond the Page time \cite{Page:1993wv,Page:2013dx}, a new saddle dominates in such a way that the radiation entropy   decreases subsequently and follows the Page curve in accordance with unitarity.

When calculating the von Neumann entropy of the subset of the Hawking radiation $R$, the new saddles are determined by the island prescription 
\cite{Penington:2019kki, Almheiri:2019qdq, Engelhardt:2014gca,Engelsoy:2016xyb,Almheiri:2019psf,Penington:2019npb,Almheiri:2019yqk} which first arose in the framework of the AdS/CFT correspondence \cite{Ryu:2006bv,Hubeny:2007xt,Faulkner:2013ana,Engelhardt:2014gca}.\footnote{For related recent works on the island prescription and its implications see \cite{Geng:2021iyq,Bhattacharya:2021jrn, Kawabata:2021hac,Bousso:2021sji,Wang:2021woy,Karananas:2020fwx,Hayden:2020vyo,Basak:2020aaa,Choudhury:2020hil,Colin-Ellerin:2020mva,Goto:2020wnk,Matsuo:2020ypv,Hernandez:2020nem, Bhattacharya:2020uun, Ling:2020laa,Chen:2020hmv, Johnson:2020mwi,Chen:2020jvn,Chandrasekaran:2020qtn,Li:2020ceg,Chen:2020uac,Alishahiha:2020qza,Hashimoto:2020cas,Giddings:2020yes,Anegawa:2020ezn,Gautason:2020tmk,Chen:2020wiq,Bhattacharya:2020ymw,Chen:2019iro,Almheiri:2019hni}.} This says that the entropy is obtained by appending an arbitrary additional set of intervals $I$ on a Cauchy surface including $R$ and then computing the extremum of the `generalized entropy'
\EQ{
S_I(R)=\underset{\partial I}\ext\Big\{\sum_{\partial I}\frac{\text{Area}(\partial I)}{4G_N}+S_\text{QFT}(R\cup I)\Big\}
\ .
\label{guz2}
}
The first term here, is the contribution of the Quantum Extremal Surfaces (QES), the boundary $\partial I$ of the island. Finally the von Neumann entropy of the reduced state on $R$ is found by minimizing over all the possible saddles
\EQ{
S(R)=\min_{I} S_I(R)\ .
}
The interpretation of $S(R)$ in the semi-classical limit of a theory of gravity is a fascinating one. It is clearly not just the na\"\i ve von Neumann entropy of a subset of the Hawking radiation. It seems to involve an implicit ensemble average, although the exact meaning of the ensemble is not completely settled. On the one hand, the ensemble is thought of as an implicit facet of the semi-classical theory of gravity that involves an average over microscopic theories \cite{Penington:2019kki}. Another aspect of this idea is that replica wormholes, like other wormholes, are associated to baby universes which have long been known to involve an implicit ensemble \cite{Marolf:2020xie,Marolf:2020rpm}. On the other hand, in what seems like an alternative interpretation, the ensemble arises as in statistical mechanics as a proxy for long-time average of an ergodic dynamical system that is equilibrating \cite{Liu:2020jsv,Pollack:2020gfa,Sasieta:2021pzj,Krishnan:2021faa}. In this latter interpretation the appearance of the ensemble is entirely universal and is not special to a quantum theory of gravity. 

The island formula implies that there  are underlying correlations in the Hawking radiation that are not captured by Hawking's ``no-island" saddle $S_\emptyset(R)\equiv S_\text{QFT}(R)$. Although  the replica wormholes have been derived in the context of the near-extremal Reissner-Nordstr\"om black holes in $3+1$ dimensions, whose $s$-wave sector is captured by Jackiw-Teitelboim gravity  in $1+1$ dimensions \cite{Jackiw:1984je,Teitelboim:1983ux}, the prescription is expected to be valid for black holes in any dimension \cite{Penington:2019npb, Almheiri:2019psy}.

Black holes evaporate very slowly for most of  their life and the adiabatic approximation applies. In this approximation, which is central to the present work, it is meaningful to associate an instantaneous temperature $T(t)$ to the black hole. Specifically, the adiabatic approximation requires that the Bekenstein-Hawking entropy satisfies,
\EQ{
\SBH(t)-S_*\gg {\cal N}\ ,
\label{adi}
}
where ${\cal N}$ is the number of flavours or species of massless fields that form the radiation and $S_*$ is a possible extremal entropy of the black hole. The adiabatic approximation is an essential input into Hawking's calculation and means the radiation emitted is quasi-thermal and so the QFT entropy is approximately that of a relativistic gas in $1+1$ dimensions,\footnote{The UV cut-off contribution  to the entropy is always present but we will ignore it because it cancels out in a physically meaningful quantity like a mutual information.}
\EQ{
S_\text{QFT}(R)\approx S_\text{rad}(R)+\text{UV-cut-off}\ .
}
For instance, if the fields are bosonic and we ignore any greybody factor, then
\EQ{
S_\text{rad}(R)=\frac{\pi{\cal N}}6\int_R T\,dt\ .
}

The contribution of the present work is to apply the adiabatic limit to the island formula prescription which has been established for near extremal black holes coupled to non-gravitating radiation baths. 
This limit leads to a simple recipe for calculating the entropy of any subset of the Hawking radiation, and in particular, it allows a simple method for identifying multiple island saddles.   Whilst explicit, analytical description of island saddles for evaporating black holes has been obtained in \cite{Hollowood:2020cou,Hollowood:2020kvk} (see also \cite{Brown:2019rox}), the simple interpretation we present here, in the adiabatic limit, is novel. 

If a  saddle involves an island $I$ then the new interpretation involves the {\em mirror} $\tilde I$, under the reflection symmetry about the horizon, which is a subset of the outgoing radiation. The mirror $\tilde I$ are the `islands in the stream'. The recipe is as follows:
\begin{enumerate}[label=\protect\circled{\arabic*}]
\item Choose a subset $\partial\tilde I\subset\partial R$, with $\text{dim}(\partial R)+\text{dim}(\partial\tilde I)$ even. The elements of $\partial\tilde I$ are the endpoints of the `islands in the stream'.\footnote{For an evaporating black hole, $\partial\tilde I$ can include a point just before the collapse that forms the black hole. This is the QES of the extremal black hole or the origin for the Schwarzschild black hole.}
\item The von Neumann entropy of the reduced state on $R$ is then
\EQ{
\boxed{S(R)=\min_I\Big\{\sum_{\partial\tilde I}\SBH (\partial\tilde I)+S_\text{rad}(R\ominus\tilde  I)\Big\}\, .}
\label{ger}
}
\end{enumerate}
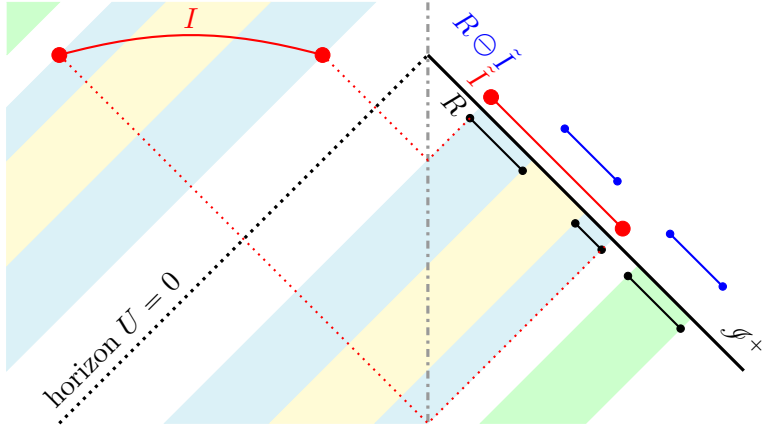
\begin{figure}[ht]
\begin{center}
\begin{tikzpicture} [scale=1.4]
\filldraw[SkyBlue!20] (2.5,1.5) -- (5.5,4.5) -- (6,4) --  (3.5,1.5) -- cycle;  
\node[rotate=-45] at (8,2.3) {\small $\mathscr I^+$};
\filldraw[yellow!20] (3.5,1.5) -- (6,4) -- (6.5,3.5) --  (4.5,1.5) -- cycle;  
\filldraw[SkyBlue!20] (5,1.5) -- (4.5,1.5) -- (6.5,3.5) -- (6.75,3.25) -- cycle;
%
\filldraw[green!20] (5.5,1.5) -- (7,3) -- (7.5,2.5) --  (6.5,1.5) -- cycle;  
%
\filldraw[SkyBlue!20] (1,2) -- (4.5,5.5) -- (3.5,5.5) --  (1,3) -- cycle;  
\filldraw[yellow!20] (1,3) -- (3.5,5.5) -- (2.5,5.5) --  (1,4) -- cycle;  
\filldraw[green!20] (1,5) -- (1.5,5.5) -- (1,5.5);  
\filldraw[SkyBlue!20] (1,4.5) -- (2,5.5) -- (2.5,5.5) --  (1,4) -- cycle;  
%
%
\draw[very thick, dotted] (1.5,1.5) -- (5,5);
\draw[very thick] (8,2) -- (5,5);
\draw[black!40,very thick, dash dot] (5,5.5) -- (5,1.5); 
\filldraw[black] (5.4,4.4) circle (1pt);
\filldraw[black] (5.9,3.9) circle (1pt);
\draw[thick] (5.4,4.4) -- (5.9,3.9);
\filldraw[black] (6.65,3.15) circle (1pt);
\filldraw[black] (6.4,3.4) circle (1pt);
\draw[thick] (6.4,3.4) -- (6.65,3.15);
\filldraw[black] (6.9,2.9) circle (1pt);
\filldraw[black] (7.4,2.4) circle (1pt);
\draw[thick] (6.9,2.9) -- (7.4,2.4);
\draw[thick,red] (5.6,4.6) -- (6.85,3.35);
\filldraw[red] (5.6,4.6) circle (2pt);
\filldraw[red] (6.85,3.35) circle (2pt);
\draw[thick,red] (1.5,5) to[out=15,in=165] (4,5);
\filldraw[red] (4,5) circle (2pt);
\filldraw[red] (1.5,5) circle (2pt);
\draw[thick,dotted,red] (5.4,4.4) -- (5,4) -- (4,5);
\draw[thick,dotted,red]  (1.5,5) -- (5,1.5) -- (6.67857,3.17857);
\node[rotate=45] at (2,2.3) {\small horizon $U=0$};
\node[red] at (2.75,5.35) {\small $I$};
\node[red,rotate=-45] at (5.5,4.8) {\small $\tilde I$};
\node[rotate=-45] at (5.25,4.55) {\small $R$};
\node [blue,rotate=-45]  at (5.55,5.15) {\small $R\ominus\tilde I$};
\filldraw[blue] (6.3,4.3) circle (1pt);
\filldraw[blue] (6.8,3.8) circle (1pt);
\draw[blue,thick] (6.3,4.3) -- (6.8,3.8);
\filldraw[blue] (7.3,3.3) circle (1pt);
\filldraw[blue] (7.8,2.8) circle (1pt);
\draw[blue,thick] (7.3,3.3) -- (7.8,2.8);
\end{tikzpicture}
\caption{\footnotesize An example where $R$ consists of 3 intervals and an island with $I$ shown as the red interval behind the horizon (actually very close to the horizon) and its mirror `island in the stream' at $\mathscr I^+$.  The coloured bands represent different subsets of outgoing modes and their entangled partner modes behind the horizon in the same colour. The light blue modes do not contribute to $S_\text{QFT}(R\cup I)$ because both out-going modes and their entangled partners are in $R\cup I$. On the other hand, the yellow and green modes do contribute because only one of the out-going modes and their partners are in $R\cup I$.}
\label{fig1} 
\end{center}
\end{figure}
The equation above involves the symmetric difference of $R$ and $\tilde I$, the union minus the intersection.\footnote{Another way to present the result is to associate the entropy of the radiation to the set of endpoints of the intervals, i.e.~$S_\text{rad}(\partial R)$. The second term in  then $S_\text{rad}(\partial R\setminus\partial\tilde I)$.} $S_{\rm BH}$ is the Bekenstein-Hawking entropy formula (arising from the area term in \eqref{guz2}) evaluated at the boundaries of $\tilde I$, which get identified with QES contributions.  Note that in the `islands in the stream' recipe the extremization of \eqref{guz2}  is implicit (in the adiabatic limit). However, to be completely clear we cannot rule out entirely the idea that there are  additional saddles, not of the type  above, that could conceivably dominate the ensemble.\footnote{There are additional saddles not of this type that appear in the analysis of \cite{Brown:2019rox}, but these never dominate the ensemble as they turn out to be  maxima.}

We illustrate the islands in the stream recipe in figure \ref{fig1}. The subset of the Hawking radiation is collected in the set of intervals $R$ near $\mathscr I^+$. In this case, the island in the stream straddles the later pair of  intervals. Note that this diagram the scales are deceptive as the island actually lies very close to the horizon. The key to the islands in the stream recipe is the relation between the outgoing null coordinates of $I$ and $R$. The outgoing modes that lie in the intersection $R\cap\tilde I$, the light blue modes, do not contribute to $S_\text{QFT}(R\cup I)$ because the islands collects the purifiers of these modes. So the modes that contribute are those that lie in the union $R\cup\tilde I$ minus the intersection, i.e.~precisely the symmetric difference. 
\begin{figure}
\begin{center}
\begin{tikzpicture} [scale=0.65]
\filldraw[fill = Plum!10!white, draw = Plum!10!white, rounded corners = 0.2cm] (-2.6,0.7) rectangle (9,-2.8);
%
\draw[decorate,very thick,black!40,decoration={snake,amplitude=0.03cm}] (0,-2.5) -- (0,0.5);
\draw[dotted,thick] (0,0) -- (8.5,0);
\draw[dotted,thick,red] (0,-1) -- (8.5,-1);
\draw[dotted,thick,blue] (0,-2) -- (8.5,-2);
\draw[very thick] (0,0) -- (5,0);
\filldraw[black] (5,0) circle (2pt);
\filldraw[black] (0,0) circle (2pt);
\draw[very thick,red] (3,-1) -- (8,-1);
\filldraw[red] (3,-1) circle (4pt);
\filldraw[red] (8,-1) circle (4pt);
\draw[very thick,blue] (0,-2) -- (3,-2);
\filldraw[blue] (0,-2) circle (2pt);
\filldraw[blue] (3,-2) circle (2pt);
\draw[very thick,blue] (5,-2) -- (8,-2);
\filldraw[blue] (5,-2) circle (2pt);
\filldraw[blue] (8,-2) circle (2pt);
\node[right] at (-2.5,0) {$R$};
\node[right] at (-2.5,-1) {$\tilde I$};
\node[right] at (-2.5,-2) {$R\ominus\tilde I$};
\draw[thick,dotted,black!60] (8,-1) -- (8,-5);
\draw[thick,dotted,black!60] (5,0) -- (5,-5);
\draw[thick,dotted,black!60] (3,-1) -- (3,-5);
\draw[thick,dotted,black!60] (0,-2) -- (0,-5);
\begin{scope}[xshift=0cm,yshift=-8cm]
\draw[thick,black!50,fill=black!20] (0,0.5) --  (8,1)  -- (8,2) -- (0,2.5) -- (-1,1.5) -- (0,0.5);
\draw[thick,black!15,fill=black!10] (8,1) -- (9,0) -- (9,3) -- (8,2);
\draw[thick,black!50,fill=black!20] (2.5,4.1) ellipse (2.5cm and 0.5cm);
%
\draw[very thick,red,dashed] (8,0.6) -- (8,3);
\draw[very thick,red,dashed]  (3,3) -- (3,0.2);
\draw[thick,green] (0.5,1.8) -- (0.5,4.2);
\draw[thick,green] (1.5,1.2) -- (1.5,3.8);
\draw[thick,green] (2.5,1.8) -- (2.5,4.2);
\draw[thick,green] (3.5,1.2) -- (3.5,3.8);
\draw[thick,green] (4.5,1.8) -- (4.5,4.2);
\draw[thick,green] (5.5,1.2) -- (5.5,3.8);
\draw[thick,green] (6.5,1.8) -- (6.5,4.2);
\draw[thick,green] (7.5,1.2) -- (7.5,3.8);
\filldraw[purple] (0.5,1.8) circle (2pt);
\filldraw[purple] (1.5,1.2) circle (2pt);
\filldraw[purple] (2.5,1.8) circle (2pt);
\filldraw[purple] (3.5,1.2) circle (2pt);
\filldraw[purple] (4.5,1.8) circle (2pt);
\filldraw[purple] (5.5,1.2) circle (2pt);
\filldraw[purple] (6.5,1.8) circle (2pt);
\filldraw[purple] (7.5,1.2) circle (2pt);
\filldraw[purple] (0.5,4.2) circle (2pt);
\filldraw[purple] (1.5,3.8) circle (2pt);
\filldraw[purple] (2.5,4.2) circle (2pt);
\filldraw[purple] (3.5,3.8) circle (2pt);
\filldraw[purple] (4.5,4.2) circle (2pt);
\filldraw[purple] (5.5,3.8) circle (2pt);
\filldraw[purple] (6.5,4.2) circle (2pt);
\filldraw[purple] (7.5,3.8) circle (2pt);
\draw[thick,blue,dashed] (0,3) -- (3,3);
\draw[thick,blue,dashed] (5,3) -- (8,3);
\filldraw[blue] (0,3) circle (2pt);
\filldraw[blue] (5,3) circle (2pt);
\filldraw[red] (3,3) circle (4pt);
\filldraw[red] (8,3) circle (4pt);
\node[right] at (-0.9,4.1) {$R$};
\end{scope}
\end{tikzpicture}
\end{center}
\caption{\footnotesize The relation between the `islands in the  stream' and the `bridge to nowhere'  \cite{Maldacena:2013xja,Brown:2019rox}. As the black hole evaporates, the bridge grows to the right and modes along the bridge are entangled with modes in the radiation as shown by the green lines. The island in the stream $\tilde I$ cancels out part of the entanglement  of $R$ with the bridge but adds in additional entanglement shown as the blue cut on the right. The contributions from the QES correspond to cutting across the `bridge to nowhere' along the red cuts. The configuration shown here will not be an extremum of the generalized entropy.}
\label{fig2} 
\end{figure}
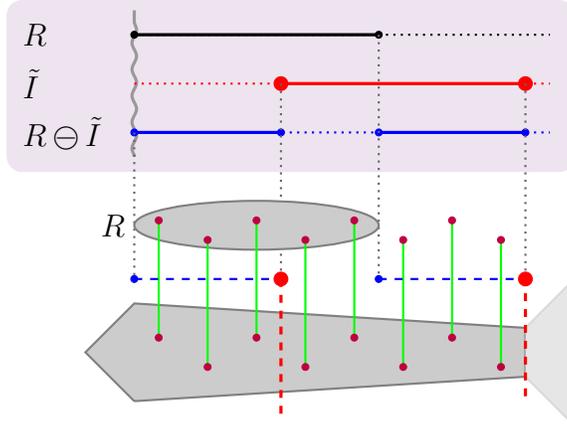

Another way to visualize the islands in the stream recipe is in relation to the `bridge to nowhere' of an evaporating black hole \cite{Brown:2019rox} as in figure \ref{fig2}. The intervals $R$, $\tilde I$ and the symmetric difference $R\ominus\tilde I$ line up with the bridge to nowhere as shown. The subset $R\ominus\tilde I$ is the blue cut between the bridge and the radiation whose entanglement entropy is $S_\text{rad}(R\ominus\tilde  I)$. The contribution from the QES corresponds to the area of the  bridge along the red dotted lines. The configuration shown in figure \ref{fig2} will not be an extremum of the generalized entropy because the endpoints of the island in the stream $\tilde I$ do not match up with the endpoints of $R$.

The paper is organized as follows. In section \ref{s2} we review important details of evaporating black holes and the adiabatic limit, and Hawking's calculation. We end with important observations about the entanglement of modes across the horizon. In section \ref{s3}, we motivate the `islands in the stream' formalism although the detailed calculation and proof of \eqref{ger} is relegated to the appendix \ref{a1}. This work is a new interpretation of the original analysis in \cite{Hollowood:2020kvk}. In section \ref{s4}, we use the formalism to describe various quantum information properties of the radiation, including its purity (i.e.~unitarity), mutual information of two intervals and then various quantum information constraints on the mutual information, including the Araki-Lieb inequality, subadditivity and strong subadditivity. In section \ref{s5} we summarize our findings and draw some conclusions.

\section{Evaporating black holes}\label{s2}

As a black hole radiates it loses mass, changing the spacetime geometry. Obtaining the backreaction on the metric by the outgoing Hawking radiation and compensating ingoing negative energy flux is complicated, requiring systematic approximations.  A particularly useful simplifying scenario emerges when restricting to the $s$-wave sector of the near horizon geometry of a near-extremal charged black hole in $3+1$ dimensions, which is captured by the black hole in JT gravity in $1+1$ dimensions. In this scenario, the JT gravity metric is fixed to be AdS$_2$ and the asymptotically flat region required to model black hole evaporation is included by gluing on a half Minkowski space to the  regularized boundary of AdS$_2$ 
\cite{Almheiri:2019psf,Almheiri:2019yqk,Almheiri:2019qdq}. Importantly, gravity is taken to be dynamical only in the AdS$_2$ region.  Within this JT gravity setup,  the fully back-reacted evaporating black hole can be solved for exactly even without invoking the adiabatic limit \cite{Almheiri:2019psf,Hollowood:2020cou}. The analysis we present in this paper is based on the generalised entropy formula \eqref{guz2} which has been derived in the effective JT gravity setup coupled to non-gravitating Minkowski radiation baths.

In the adiabatic limit, however, as we will see below  we are able to  keep the discussion fairly general 
where possible, although the specific computations follow from the effective 1+1 dimensional picture described above. 

It is useful at this stage to briefly review some of the issues surrounding the applicability of the island picture when the bath regions gravitate. For the situation with dynamical gravity in the bath, we adopt the viewpoint advocated in the work of  Marolf  and Maxfield \cite{Marolf:2020rpm} wherein replica wormholes are shown to contribute to measurements of R\'enyi entropies (realized as `swap' entropies) of subsets of Hawking radiation, performed  by asymptotic observers with fixed geometry at $\mathscr{I}^+$ (while the geometry is dynamical everywhere else). Within the semi-classical approximation, these entropies are calculated as saddle points of the functional integral where the replicas and their conjugates are sewn together along a Cauchy surface that asymptotes to the relevant portion of $\mathscr I^+$ but avoids regions of large curvature. This picture naturally leads to a corresponding QES/island prescription in asymptotically flat space and demonstrates that the entropy in the semi-classical approximation $S(R)$ is actually an average over an ensemble associated to a Hilbert space of baby universe states on the island. 

We  note that  the QES/island prescription and indeed the Page curve itself when gravity is dynamical in the bath region, has been questioned in recent works, both on general grounds \cite{Raju:2020smc, Laddha:2020kvp} and within the framework of Karch-Randall braneworlds scenario \cite{Geng:2020fxl, Geng:2020qvw, Geng:2021hlu} \footnote{We thank Suvrat Raju for correspondence on this and related issues.}.  Our view however, is that the setup involving non-gravitating baths allows a crisp definition of the generalized entropy which unambiguously exhibits a Page curve. When the bath is made dynamical,  suitably defined  coarse-grained entropies in the (semiclassical) gravitating effective field theory can then be expected to display the same behaviour (see e.g. \cite{Ghosh:2021axl}).

\subsection{Mass and entropy in the adiabatic limit}

Hawking's calculation proceeds by calculating the occupation number of outgoing modes of frequency $\omega$ in the asymptotically flat region far from the hole, and yields a result \cite{Hawking:1974sw,Hawking:1976ra} (summarized in the textbook \cite{BD})
\EQ{
\bar N_\omega=\frac{\Gamma(\omega)}{e^{\omega/T}-1}\ ,
\label{kex}
}
where $T$ is the temperature of the black hole when the mode leaves the vicinity of the horizon. We will assume that the Hawking modes are associated to a large number $\mathcal{N}$ of massless scalars, in order to justify the  semi-classical approximation, and that most of the evaporation occurs via the $s$-wave mode so the outgoing modes are effectively a $1+1$-dimensional relativistic bosonic gas. Adding in the higher angular momentum modes is straightforward. In the above, $\Gamma(\omega)$, is the greybody factor that accounts for the tunneling probability for modes through the effective potential around the black hole. In the following, for simplicity, we shall ignore this effect and set $\Gamma(\omega)=1$. In that case, eq.~\eqref{kex} is the Planck spectrum.  The calculation is valid in the adiabatic limit where the evaporation is slow enough that the back reaction of the emitted Hawking radiation is simply taken into account by associating a slowly varying, time-dependent temperature. 

The entropy of the radiation in a thin shell of width $dt$ is then given by that of a $1$-dimensional relativistic bosonic gas of volume $dt$:
\EQ{
dS_\text{rad}=dt\,{\cal N}\int_0^\infty\frac{d\omega}{2\pi}\big((\bar N_\omega+1)\log(\bar N_\omega+1)-\bar N_\omega\log\bar N_\omega\big)=\frac{\pi {\cal N}T}6\,dt\,.
\label{hxx}
}
Therefore  the entropy of the radiation emitted between the formation of the black hole at $t=0$ and a later time $t$ is
\EQ{
S_\text{rad}(t)=\frac{\pi {\cal N}}6\int_0^tT\,dt\ .
\label{dup}
}

The temperature $T$ is determined by the energy conservation equation which equates the rate of change of the mass with minus  the outgoing flux of energy of the Hawking radiation 
\EQ{
\dot M=-{\cal N}\int_0^\infty\frac{d\omega}{2\pi}\cdot\frac{\omega}{e^{\omega/T}-1}=-\frac{\pi{\cal N}T^2}{12}\ ,
\label{her}
}
where ${\cal N}$ are the number of massless scalar fields. In order to solve this, it is necessary to know the relation between the mass $M$ and the temperature $T$ of the black hole. This relation depends on the  specific black hole solution. 

For example, for the black hole in JT gravity \cite{Jackiw:1984je, Teitelboim:1983ux} which yields a good approximation to the dynamics of the near horizon AdS$_2$ factor of a 
near-extremal charged black hole, the mass
\EQ{
M=M_*+\frac{\pi\phi_rT^2}{4G_N}\,,
}
where $M_*$ is the mass of the extremal black hole. Formally, within the JT gravity framework, in order to describe black hole evaporation with an outgoing radiation flux, we have to implicitly assume transparent boundary conditions at the boundary of the AdS$_2$. Solving equation \eqref{her} then gives
\EQ{
T(t)=T_0e^{-kt/2}\ ,
\label{bot}
}
where
\EQ{
k=\frac{{\cal N}G_N}{3\phi_r}\ ,
\label{sew}
}
In the above, $\phi_r$ sets the boundary value of the dilaton. When the JT gravity descends from the near extremal charged black hole in $3+1$ dimensions, this is set to be $\phi_r=8\pi(G_NM_*)^3$. 

On the other hand, for the Schwarzschild black hole in 3+1 dimensions, the mass is inversely proportional to the temperature:
\EQ{
M=\frac1{8\pi G_NT}
}
and so solving \eqref{her}, we have
\EQ{
T(t)=T_0(1-t/t_\text{evap.})^{-1/3}\ ,
\label{bat}
}
where the evaporation time is
\EQ{
t_\text{evap.}=\frac{2^8\pi G_N^2M_0^3}{\cal N} .
\label{fit}
}

The Bekenstein-Hawking entropy of the evaporating black hole can be obtained by integrating up the thermodynamic relation $d\SBH =dM/T$ along with \eqref{her},
\EQ{
\SBH(t)=\frac{\text{Area}(\text{horizon})}{4G_N}=S_*+\frac{\pi{\cal N}}{12}\int^{t_\text{evap.}}_t T\,dt\ ,
\label{leg}
}
where $t_\text{evap.}$ is the endpoint of the evaporation; notice that $t_\text{evap.} = \infty$ for the near-extremal Reissner-Nordstr\"om black hole or black hole in JT gravity. $S_*$ is the extremal entropy which vanishes for the Schwarzschild black hole. The Bekenstein-Hawking entropy 
\EQ{
\ZZ\equiv\SBH(t)\ ,\label{zz}
} 
provides a universal time coordinate to describe the decay. This coordinate decreases from $\ZZ_0\equiv \SBH (0)$ the initial entropy of the black hole just after it has formed. It is worth noting that in the adiabtaic limit and without greybody factors, the Bekenstein-Hawking entropy \eqref{leg} and entropy of the radiation \eqref{dup} are simply related,
\EQ{
S_\text{rad}(t)=2(\ZZ_0-\ZZ)\ ,
}
so the  radiation carries away twice the entropy decrease of the  black hole.\footnote{This  relation is changed by including a non-trivial greybody factor \cite{Page:1993wv,Page:2013dx}.}

The adiabatic approximation is a key part of Hawking's calculation. Intuitively, it is  in the limit of slow evaporation  where  the temperature is slowly changing, that  the radiation can be described by equilibrium formulae  \eqref{hxx} and  \eqref{her} that depend implicitly on  time via the temperature. The adiabatic limit is valid when \eqref{adi} is satisfied and it breaks down near the end of the evaporation. Note that we are also working in the semi-classical limit ${\cal N}\gg1$. In addition, we require that any subsets of the  Hawking radiation $R$ we focus attention on, must be sufficiently large that their  von Neumann entropy in the QFT is captured by the thermodynamic entropy \eqref{dup} i.e.~we must always work in the thermodynamic limit. This requires that the subset endpoints $u_i,u_j\in\partial R$, $u_i<u_j$, are well separated  so that,
\EQ{
	S_\text{rad}\left([u_i,u_j]\right)\gg{\cal N}\ .
	\label{ab2}
} 

\subsection{Coordinates and the generation of Hawking radiation}

At its heart the generation of Hawking radiation is the tale of two coordinate frames. The first $(U,V)$ are  Kruskal Szekeres (KS) coordinates. These naturally cover both inside and outside the horizon. In JT gravity this is simply the metric of AdS$_2$,
\EQ{
ds^2=-\frac{dU\,dV}{(1+UV)^2}\ ,
\label{zik}
}
reflecting the fact that the near-horizon geometry of the near-extremal charged black hole in $3+1$ dimensions is AdS$_2\times S^2$. The temperature of the black hole and its mass are inferred from the dilaton. The horizon is $U=0$ and in its vicinity $ds^2\approx -dU\,dV$, therefore, in this region  the frame $(U,V)$ is inertial. The second frame is $(u,v)$, the Schwarzschild coordinates, that describe a non-inertial frame at fixed $r$ from the black hole. However, the situation flips a long way from the black hole: there $(U,V)$ become non-inertial, while the coordinates $(u,v)$ become the inertial, the null coordinates of the asymptotically flat region  $ds^2\approx -du\,dv$. In the JT gravity set up $(u,v)$ are null coordinates in the Minkowski region that is glued onto the boundary of $\text{AdS}_2$.

We will write the coordinate transformation between the frames as 
\EQ{
U=-\exp\Big[-2\pi\int^u T\,dt\Big]\ ,\qquad V=\exp\Big[2\pi\int^v T\,dt\Big]\ ,
\label{pop}
}
for a function $T(t)$. In the adiabatic limit, we will identify $T(t)$ as the instantaneous temperature of the black hole. This function completely determines the exact back-reacted solution in JT  gravity because the metric is fixed to be \eqref{zik} and all the non-trivial back-reaction effects are on the dilaton whose solution can be found exactly \cite{Hollowood:2020cou}:\footnote{In \cite{Hollowood:2020cou}, we used the  notation $(w^+,w^-)=(V,U)$ and $(y^+,y^-)=(v,u)$ and  defined the function $\hat f(t)$ so that $w^\pm=\pm\hat f(y^\pm)^{\pm1}$, so that $\hat f(t)=\exp2\pi\int^t T\,dt$.}
\EQ{
\phi=\phi_0+2\pi\phi_r T(v)\frac{1-UV}{1+UV}+\phi_r\frac{\dot T(v)}{T(v)}\ .
\label{dil}
}
The third term here can be ignored in the adiabatic approximation. In this limit and in the near-horizon region, that is $|UV|\ll1$, we can write
\EQ{
\frac{\phi}{4G_N}\Big|_\text{near hor.}\equiv \frac{\text{Area}(S^2)}{4G_N}=S_*+(\SBH (v)-S_*)(1-2UV)+\cdots\ ,
\label{nil}
}
We have identified the dilaton with the area of the  $S^2$ of the  Reissner-Nordstr\"om black hole. This expression also applies to the Schwarzschild black hole in the adiabatic limit where the metric takes the form of the Vaidya metric.

\subsection{The quantum state}

A key step in the theory of Hawking radiation is the specification of the quantum state of the QFT near the horizon.\footnote{In the original calculation Hawking specified the quantum state on $\mathscr I^-$ rather than near the horizon, however, Jacobson \cite{Jacobson:2003vx}  showed that it was equivalent to work in the near-horizon region.}
For an evaporating black hole, the appropriate state is the Unruh state for which the outgoing modes are in the vacuum state of the KS frame associated to  the $U$ coordinate, and the infalling modes are in the vacuum state of the Schwarzschild coordinates associated to the $v$ coordinate. The derivation of Hawking radiation, e.g.~\cite{BD}, follows by calculating the expectation value of the occupation number $\bar N_\omega$ of an outgoing positive frequency mode $Z_\omega$ of frequency $\omega$ in the $u$ frame in the asymptotically flat regime. The calculation is an example of an in-in type calculation and proceeds by tracing  the mode $Z_\omega$ backwards in time till it is near the horizon. This generally involve a tunneling  probability, i.e.~greybody factor, but we are ignoring that here. Near the horizon, the quantum state of the outgoing modes is the $U$ vacuum and the mode $Z_\omega$ becomes a mixture of a positive and negative frequency mode in the $U$ frame. The latter contribution means that the $U$ vacuum contains a non-vanishing occupation number of $Z_\omega$ modes. The occupation number $\bar N_\omega$ in \eqref{kex}, with $\Gamma=1$, follows from a Bogoliubov transformation.

\subsection{Stress tensor}

A simpler way to calculate the flux of Hawking radiation is to consider the stress tensor of the outgoing modes \cite{BD}. In the  Unruh state, i.e.~$U$ vacuum, we have $T_{UU}=0$.\footnote{These expressions are renormalized expectation values in the semi-classical limit.} We can find the outgoing flux by making the conformal transformation $U\to u$. The stress tensor picks up a contribution from the conformal anomaly that involves the Schwarzian derivative that we can write  in terms of the temperature $T(u)$:
\EQ{
T_{uu}=-\frac{\cal N}{24\pi}\{U,u\}=\frac{\pi {\cal N}}{12}\Big(T^2+\frac3{4\pi^2}\Big(\frac{\dot T}{T}\Big)^2-\frac{\ddot T}{2\pi^2T}\Big)\ .
}

We recognize the first term here as the thermodynamic energy flux of the boson gas in \eqref{her}. Hence, the adiabatic approximation is valid when the first term dominates:
\EQ{
\Big|\frac{\dot T^2}{T^4}-\frac{2\ddot T}{3T^3}\Big|\ll1\ ,
\label{sup}
}
and in this case we get $T_{uu}=\pi{\cal N}T^2/12$, which is the Stefan-Boltzmann law, i.e. the outgoing energy flux of a relativistic bosonic gas of temperature $T$ in \eqref{her}. 

Note that in the Unruh state, the infalling modes are in the $v$ vacuum and so  $T_{vv}=0$. This implies that  in this case 
\EQ{
T_{VV}=\frac{\cal N}{24\pi}\Big(\frac{\partial v}{\partial V}\Big)^2\{V,v\}=\Big(\frac{\partial v}{\partial V}\Big)^2\dot M(v)<0
}
and so there is a negative energy flux into the black hole.  This is of course entirely consistent because the energy carried by the outgoing Hawking modes must be balanced by an infalling negative energy flux that reduces the mass of the black hole.

In the JT gravity setup, we can proceed further without invoking the adiabatic limit \cite{Almheiri:2019psf,Hollowood:2020cou}. The mass of the black hole $M=M_*+E$ where 
\EQ{
E(t)=-\frac{\phi_r}{8\pi G_N}\{V,v\}\Big|_{v=t}
\label{yet}
}
is the ADM mass in JT gravity, and note that $\{V,v\}_{v=t}=\{U,u\}_{u=t}$. Hence, the energy conservation equation
\EQ{
\dot M=-T_{uu}\Big|_{u=t}\ ,
}
is simply
\EQ{
\dot E=-kE\ ,
}
where $k$ has been defined in \eqref{sew}. Of course, this implies an exponential decay of energy with time,
\EQ{
E(t)=E_0e^{-kt}\ .
}
The function $V(v)$ can then be found by solving \eqref{yet}. The solution is known exactly in terms of Bessel functions \cite{Almheiri:2019psf,Hollowood:2020cou}.

However, if we \emph{do} invoke the adiabatic limit, then the energy conservation equation can be written as a simple equation for the temperature
\EQ{
\dot T=-\frac{kT}2\ ,
}
whose solution is \eqref{bot}. We can check the requirement for the adiabatic approximation \eqref{sup} in this case,
\EQ{
\frac{k^2e^{kt}}{8T_0^2}\ll 1
}
and so the adiabatic approximation is valid until the very long time scale
\EQ{
k^{-1}\log(T_0/k)\thicksim k^{-1}\log\frac{\SBH (0)-S_*}{\cal N}\ .
}
For the Schwarzschild black hole, a similar analysis shows that the adiabatic approximation is valid until near the end of the evaporation up to time scale $1-t/t_\text{evap.}\sim(G_NM_0)^{-3/2}$ when the black hole is Planck sized. 

\subsection{QFT and thermodynamic entropies}

In the adiabatic limit, the QFT entropy of the reduced state on an interval of Hawking modes $R=[u_1,u_2]$ near $\mathscr I^+$ has a simple limit when the interval is large in the sense of \eqref{ab2} which implies that the relative magnitude of the KS coordinates, 
\EQ{
U_1/U_2\gg1\ .
\label{fot}
}
This is roughly the limit for which $\Delta u\gg T^{-1}$. In this limit, and ignoring the cut off term,
\EQ{
S_\text{QFT}(R)=\frac{\cal N}6\log\frac{U_2-U_1}{\sqrt{U_1U_2}}\approx\frac{\cal N}6\log\sqrt{\frac{U_1}{U_2}}=S_\text{rad}(R)\ .
\label{gcx}
}
In terms of the $\ZZ$ coordinate \eqref{zz}, 
\EQ{
S_\text{QFT}(R)\approx2(\ZZ_1-\ZZ_2)\ .
}

\subsection{Purifiers}

The outgoing modes are in the $U$ vacuum state. This state has a non-vanishing occupation number of positive frequency modes ${\cal Z}_\omega$ with respect to the $u$ vacuum which is the Minkowski vacuum far from the black hole. These modes ${\cal Z}_\omega(U)$ have support outside the horizon $U<0$. Each has a partner mode 
\EQ{
\tilde {\cal Z}_\omega(U)={\cal Z}_\omega(-U)\ ,
}
with support behind the horizon $U>0$. The pair of modes are entangled in the $U$ vacuum when considered relative to the $u$ vacuum,
\EQ{
\ket{0}_U\thicksim \prod_\omega\exp\Big[e^{-\omega/2T}a^\dagger({\cal Z}_\omega)a^\dagger(\tilde {\cal Z}^*_\omega)\Big]\ket{0}_u\ .
}
The entropy of the outgoing Hawking radiation is precisely the entanglement entropy obtained by tracing out the partner modes inside the horizon. 

The modes ${\cal Z}_\omega$ above have definite  momentum and so are completely de-localized in $u$. However, we can make  localized wave packets  by smearing with some $\Delta\omega$. Since the characteristic frequency is $\sim T^{-1}$, according to the uncertainty principle we should able to define localized modes with $\Delta u\sim 1/\Delta\omega\sim T^{-1}$. This intuition is important, because it means that an interval of Hawking modes $R=[u_1,u_2]$ with $\Delta u\gg T^{-1}$ will be entangled with a definite interval of modes behind the horizon related by the symmetry $U\to-U$. 
\begin{figure}[ht]
\begin{center}
\begin{tikzpicture} [scale=1]
\filldraw[SkyBlue!20] (4,0) -- (7,3) -- (6,4) --  (2,0) -- cycle;  
\filldraw[red!20] (0,3) -- (0,1) -- (4,5) --  (2,5) -- cycle;  
\draw[very thick, dotted] (0,0) -- (5,5);
\draw[very thick] (8,2) -- (5,5);
\draw[black!40,very thick, dash dot] (5,5) -- (5,0); 
\draw[black!40,thick,->] (5,0) -- (0.3,4.7);
\node at (0.1,4.9) {$U$};
\node[rotate=-45] at (4.3,0.3) {$U_1$};
\node[rotate=-45] at (3.3,1.3) {$U_2$};
\node[rotate=-45] at (0.8,3.8) {$U_4$};
\node[rotate=-45] at (1.8,2.8) {$U_3$};
\node[rotate=-45] at (3.5,0.5) {$R$};
\node[rotate=-45] at (1,3) {$I$};
\node[rotate=45] at (4,4.3) {horizon};
\node[rotate=45] at (0.9,1.2) {$U=0$};
\node[rotate=-45] at (6.7,3.7) {${\mathscr  I}^+$};
\end{tikzpicture}
\caption{\footnotesize A subset of outgoing modes in the interval $R=[U_1,U_2]$ at $\mathscr I^+$ and an interval $I=[U_3,U_4]$ behind the horizon (so with an order $1,2,3,4$ along a Cauchy slice that asymptotes to $\mathscr I^+$ from right to left).}
\label{fig3} 
\end{center}
\end{figure}
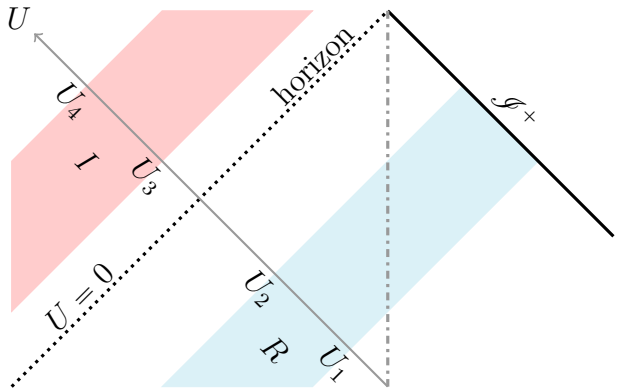
In order to show this quantitatively, consider the modes in an interval $R$ at $\mathscr I^+$ and an interval $I$ behind the horizon shown in figure \ref{fig3}. We are assuming that the interval $I$ is in the near-horizon region $UV\ll1$. Let us suppose that,
\EQ{
|U_1|\gg U_4\gg|U_2|\gg U_3
\label{vef}
}
and that the differences between these coordinates is  sufficiently large that the entropy of the corresponding intervals in the $U$ vacuum is captured by the thermodynamic entropy. The question is what is $S_\text{QFT}(R\cup I)$? 

The answer in the QFT can be computed because the theory is free. The contribution from the infalling modes is subleading and will be ignored, hence\footnote{There is a subtlety here. The Hawking radiation in interval $R$ is in the vacuum associated to the $U$ frame and the conformal factors for $u\to U$ contribute the factors $\sqrt{U_1U_2}$ below. However, for the interval $I$ in the near-horizon region there are no such conformal factors because the metric here is inertial in the $U$ frame, $ds^2\approx-dU\,dV$. But serendipitously it turns out that the factors of $\sqrt{U_3U_4}$ arise from the infalling mode sector because the endpoints of  $I$ will have coordinates related by $U\sim V^{-1}$.}
\begin{equation}
S_\text{QFT}(R\cup I)=\frac{\cal N}6\log\frac{U_2-U_1}{\sqrt{U_1U_2}}+\frac{\cal N}6\log\frac{U_4-U_3}{\sqrt{U_3U_4}}
+\frac{\cal N}6\log\frac{(U_4-U_1)(U_3-U_2)}{(U_4-U_2)(U_3-U_1)}\ .
\end{equation}
Now we take the limit \eqref{vef},
\EQ{
S_\text{QFT}(R\cup I)&\approx 
\frac{\cal N}{12}\log\frac{-U_2}{U_3}+\frac{\cal N}{12}\log\frac{-U_1}{U_4}\\[5pt]
&=S_\text{rad}([u_2,\tilde u_3])+S_\text{rad}([u_1,\tilde u_4])\ .
}
Here, we have introduced a coordinate $\tilde u$ behind the horizon where
\EQ{
U=\exp\Big[-2\pi\int^{\tilde u}T\,dt\Big]\,.
}
(Compare this with the relation between $U$ and $u$ outside the horizon, in \eqref{pop}). However, we can also think of $\tilde u_i$ as the outgoing null coordinate associated to the mirror of the endpoints of $I$, i.e.~the map $U\to-U$ which sends $\tilde u\to u=\tilde u$.

The simple result here is significant. The interpretation is that modes in $[\tilde u_4,u_2]\subset R$ have their purifiers inside $I$ and so do not contribute to the entropy of $R\cup I$. On the other hand the modes in the interval $[u_2,\tilde u_3]$ add to the entropy. One can see that the end result can be written using the symmetric difference of set theory
\EQ{
S_\text{QFT}(R\cup I)\overset{\text{adiabatic}}=S_\text{rad}(R\ominus\tilde I)\ ,
\label{vzz}
}
where $\tilde I$ is the mirror image of the interval $I$ at $\mathscr I^+$, under the symmetry $U\to-U$. When $I$ is an island, then its mirror $\tilde I$ at $\mathscr I^+$ is the  `island in the stream'. We can visualize the overlapping of the sets of modes as follows:
{\small
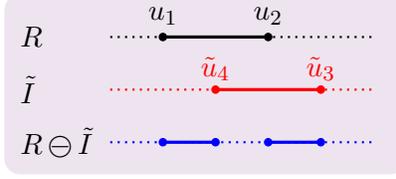
\begin{center}
\begin{tikzpicture} [scale=0.7]
\filldraw[fill = Plum!10!white, draw = Plum!10!white, rounded corners = 0.2cm] (-2,0.8) rectangle (5.6,-2.6);
\draw[dotted,thick] (0,0) -- (5,0);
\draw[dotted,thick,red] (0,-1) -- (5,-1);
\draw[very thick] (1,0) -- (3,0);
\filldraw[black] (1,0) circle (2pt);
\filldraw[black] (3,0) circle (2pt);
\draw[very thick,red] (2,-1) -- (4,-1);
\filldraw[red] (2,-1) circle (2pt);
\filldraw[red] (4,-1) circle (2pt);
\draw[dotted,thick,blue] (0,-2) -- (5,-2);
\draw[very thick,blue] (1,-2) -- (2,-2);
\draw[very thick,blue] (3,-2) -- (4,-2);
\filldraw[blue] (1,-2) circle (2pt);
\filldraw[blue] (2,-2) circle (2pt);
\filldraw[blue] (3,-2) circle (2pt);
\filldraw[blue] (4,-2) circle (2pt);
\node[right] at (-1.9,-2) {$R\ominus\tilde I$};
\node[right] at (-1.9,0) {$R$};
\node[right] at (-1.9,-1) {$\tilde I$};
\node at (1,0.4) {$u_1$};
\node[red] at (2,-0.6) {$\tilde u_4$};
\node at (3,0.4) {$u_2$};
\node[red] at (4,-0.6) {$\tilde u_3$};
\end{tikzpicture}
\end{center}}
Note that if we take $\tilde I$ to be $R$ then 
\EQ{
S_\text{QFT}(R\cup \tilde R)=0\ ,
}
to leading order in the adiabatic approximation
and so the intervals $\tilde R$ and $R$ are mutual purifiers.

\section{Islands in the stream}\label{s3}

In this section we infer the existence of a class of extrema of the generalized entropy using intuitive arguments. The detailed proof is relegated to appendix \ref{a1}. 

The first assumption is that the QES lie close to the horizon. We can show that this is self consistent by  the following argument. The infalling modes are in the $v$ vacuum and their contribution to the QFT entropy at leading order is only via the conformal factor in the AdS region. Hence, given \eqref{nil}, the dependence of the generalized entropy on each of the QES coordinates $V_a$ is particularly simple
\EQ{
S_\text{gen.}=(\SBH (v_a)-S_*)(1-2U_aV_a)+\frac{\cal N}{12}\log V_a+\cdots\ .
\label{cco}
}
Then using  the fact that
\EQ{
\frac{d\SBH }{dV}=\frac1T\cdot\frac{dM}{dV}=-\frac{\cal N}{24V}\  ,
\label{nip2}
}
the extremization of the generalized entropy respect to $V_a$ gives
\EQ{
U_aV_a=\frac{\cal N}{48(\SBH (v_a)-S_*)}\ ,
\label{hik2}
}
and so in the adiabatic limit \eqref{adi}, $U_aV_a\ll1$ and the QES is, indeed, close to the horizon. 

 There is a subtle point regarding \eqref{hik2}. It ensures that the QES are close to the horizon, however, it is actually a subleading effect because the leading order behaviour cancels out in the combination $U_aV_a$. As we will see, this subleading term is responsible for the scrambling time, while at leading order we have  $\tilde u_a = v_a$. 
So the contribution to the generalized entropy \eqref{cco} can be written as the Bekenstein-Hawking entropy evaluated at the time $\tilde u_a$. In particular, this is the outgoing coordinate of the mirror of the QES $\partial\tilde I$ under the symmetry $U\to-U$. If we insert this into the generalized entropy and use \eqref{vzz}, then at leading order we have
\EQ{
S_\text{gen.}=\sum_{\partial\tilde I}\SBH (\partial\tilde I)+S_\text{rad}(R\ominus\tilde I)\ .
}
The remaining task is to extremize over the outgoing coordinates of the QES, i.e.~the endpoints of the islands in the stream $\tilde I$ with coordinates $\tilde u_a$.

Consider the variation of the generalized entropy as one of the endpoints of $\tilde I$ varies in the neighborhood of an endpoint of $\partial R$. There are four possible scenarios, the first two with $\tilde u_a$ increasing through $u_i$ from left to right, are
{\small
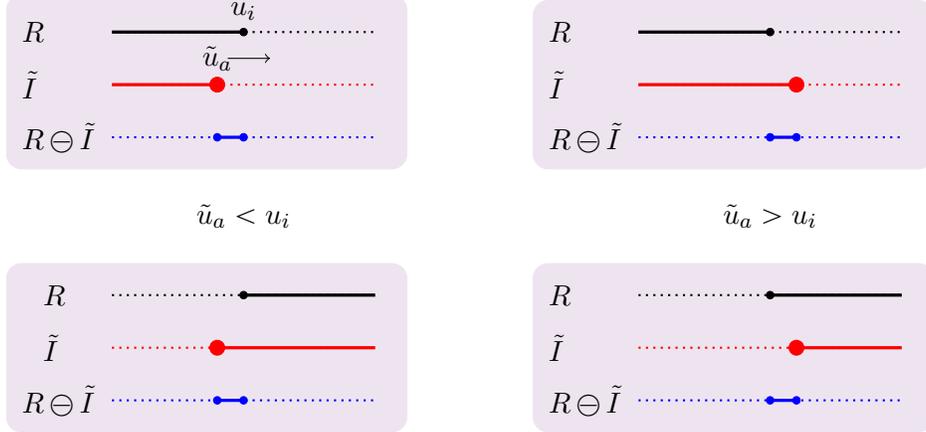
\begin{center}
\begin{tikzpicture} [scale=0.7]
\filldraw[fill = Plum!10!white, draw = Plum!10!white, rounded corners = 0.2cm] (-2,0.7) rectangle (5.6,-2.6);
\draw[dotted,thick] (0,0) -- (5,0);
\draw[dotted,thick,red] (0,-1) -- (5,-1);
\draw[very thick] (0,0) -- (2.5,0);
\filldraw[black] (2.5,0) circle (2pt);
\draw[very thick,red] (0,-1) -- (2,-1);
\filldraw[red] (2,-1) circle (4pt);
\draw[dotted,thick,blue] (0,-2) -- (5,-2);
\draw[very thick,blue] (2,-2) -- (2.5,-2);
\filldraw[blue] (2,-2) circle (2pt);
\filldraw[blue] (2.5,-2) circle (2pt);
\node[right] at (-1.9,-2) {$R\ominus\tilde I$};
\node[right] at (-1.9,0) {$R$};
\node[right] at (-1.9,-1) {$\tilde I$};
\node at (2.5,0.4) {$u_i$};
\node at (2,-0.5) {$\tilde u_a$};
\draw[->] (2.2,-0.5) -- (3,-0.5);
\node at (2.5,-3.5) {$\tilde u_a<u_i$};
\begin{scope}[xshift=10cm]
\filldraw[fill = Plum!10!white, draw = Plum!10!white, rounded corners = 0.2cm] (-2,0.6) rectangle (5.6,-2.6);
\draw[dotted,thick] (0,0) -- (5,0);
\draw[dotted,thick,red] (0,-1) -- (5,-1);
\draw[very thick] (0,0) -- (2.5,0);
\filldraw[black] (2.5,0) circle (2pt);
\draw[very thick,red] (0,-1) -- (3,-1);
\filldraw[red] (3,-1) circle (4pt);
\draw[dotted,thick,blue] (0,-2) -- (5,-2);
\draw[very thick,blue] (3,-2) -- (2.5,-2);
\filldraw[blue] (3,-2) circle (2pt);
\filldraw[blue] (2.5,-2) circle (2pt);
\node[right] at (-1.9,-2) {$R\ominus\tilde I$};
\node[right] at (-1.9,0) {$R$};
\node[right] at (-1.9,-1) {$\tilde I$};
\node at (2.5,-3.5) {$\tilde u_a>u_i$};
\end{scope}
\begin{scope}[yshift=-5cm]
\filldraw[fill = Plum!10!white, draw = Plum!10!white, rounded corners = 0.2cm] (-2,0.6) rectangle (5.6,-2.6);
\draw[dotted,thick] (0,0) -- (5,0);
\draw[dotted,thick,red] (0,-1) -- (5,-1);
\draw[very thick] (5,0) -- (2.5,0);
\filldraw[black] (2.5,0) circle (2pt);
\draw[very thick,red] (5,-1) -- (2,-1);
\filldraw[red] (2,-1) circle (4pt);
\draw[dotted,thick,blue] (0,-2) -- (5,-2);
\draw[very thick,blue] (2,-2) -- (2.5,-2);
\filldraw[blue] (2,-2) circle (2pt);
\filldraw[blue] (2.5,-2) circle (2pt);
\node[right] at (-1.9,-2) {$R\ominus\tilde I$};
\node[right] at (-1.5,0) {$R$};
\node[right] at (-1.5,-1) {$\tilde I$};
\begin{scope}[xshift=10cm]
\filldraw[fill = Plum!10!white, draw = Plum!10!white, rounded corners = 0.2cm] (-2,0.6) rectangle (5.6,-2.6);
\draw[dotted,thick] (0,0) -- (5,0);
\draw[dotted,thick,red] (0,-1) -- (5,-1);
\draw[very thick] (5,0) -- (2.5,0);
\filldraw[black] (2.5,0) circle (2pt);
\draw[very thick,red] (5,-1) -- (3,-1);
\filldraw[red] (3,-1) circle (4pt);
\draw[dotted,thick,blue] (0,-2) -- (5,-2);
\draw[very thick,blue] (3,-2) -- (2.5,-2);
\filldraw[blue] (3,-2) circle (2pt);
\filldraw[blue] (2.5,-2) circle (2pt);
\node[right] at (-1.9,-2) {$R\ominus\tilde I$};
\node[right] at (-1.9,0) {$R$};
\node[right] at (-1.9,-1) {$\tilde I$};
\end{scope}
\end{scope}
\end{tikzpicture}
\end{center}}
\noindent The QES contributes $\ZZ_a\equiv \SBH (\tilde u_a)$ to the entropy while the contribution to $S_\text{QFT}(R\ominus\tilde I)$ comes from the blue region is $S_\text{rad}=2|\ZZ_a-\ZZ_i|$. Hence as $\tilde u_a$ increases, we have
\EQ{
\frac{\partial S_\text{gen.}}{\partial\tilde u_a}=\begin{cases} 3\dot{\EuScript S}_a & \tilde u_a<u_i\ ,\\ -\dot{\EuScript S}_a& \tilde u_a>u_i\ ,\end{cases}
}
for either of the two scenarios. Since $\dot{\EuScript S}<0$, the entropy has a minimum at $\tilde u_a=u_i$. 

The other two possible scenarios are
{\small
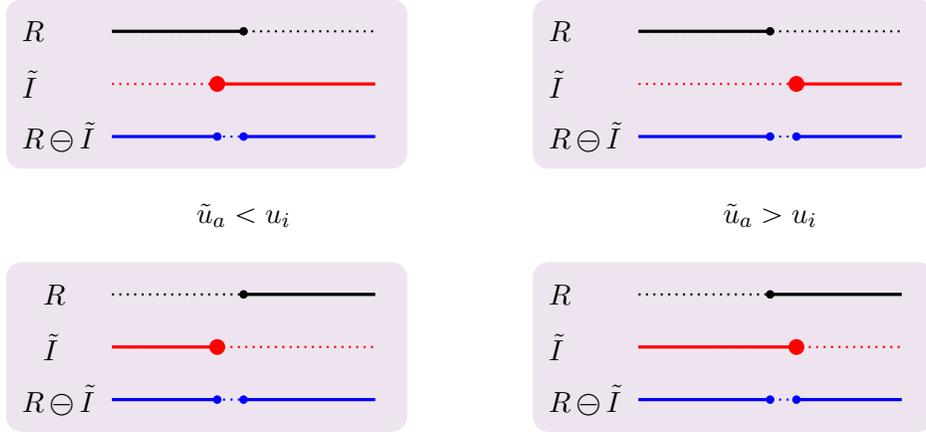
\begin{center}
\begin{tikzpicture} [scale=0.7]
\filldraw[fill = Plum!10!white, draw = Plum!10!white, rounded corners = 0.2cm] (-2,0.6) rectangle (5.6,-2.6);
\node at (2.5,-3.5) {$\tilde u_a<u_i$};
\draw[dotted,thick] (0,0) -- (5,0);
\draw[dotted,thick,red] (0,-1) -- (5,-1);
\draw[very thick] (0,0) -- (2.5,0);
\filldraw[black] (2.5,0) circle (2pt);
\draw[very thick,red] (5,-1) -- (2,-1);
\filldraw[red] (2,-1) circle (4pt);
\draw[dotted,thick,blue] (0,-2) -- (5,-2);
\draw[very thick,blue] (0,-2) -- (2,-2);
\draw[very thick,blue] (2.5,-2) -- (5,-2);
\filldraw[blue] (2,-2) circle (2pt);
\filldraw[blue] (2.5,-2) circle (2pt);
\node[right] at (-1.9,-2) {$R\ominus\tilde I$};
\node[right] at (-1.9,0) {$R$};
\node[right] at (-1.9,-1) {$\tilde I$};
\begin{scope}[xshift=10cm]
\filldraw[fill = Plum!10!white, draw = Plum!10!white, rounded corners = 0.2cm] (-2,0.6) rectangle (5.6,-2.6);
\draw[dotted,thick] (0,0) -- (5,0);
\draw[dotted,thick,red] (0,-1) -- (5,-1);
\draw[very thick] (0,0) -- (2.5,0);
\filldraw[black] (2.5,0) circle (2pt);
\draw[very thick,red] (5,-1) -- (3,-1);
\filldraw[red] (3,-1) circle (4pt);
\draw[dotted,thick,blue] (0,-2) -- (5,-2);
\draw[very thick,blue] (0,-2) -- (2.5,-2);
\draw[very thick,blue] (3,-2) -- (5,-2);
\filldraw[blue] (3,-2) circle (2pt);
\filldraw[blue] (2.5,-2) circle (2pt);
\node[right] at (-1.9,-2) {$R\ominus\tilde I$};
\node[right] at (-1.9,0) {$R$};
\node[right] at (-1.9,-1) {$\tilde I$};
\node at (2.5,-3.5) {$\tilde u_a>u_i$};
\end{scope}
\begin{scope}[yshift=-5cm]
\filldraw[fill = Plum!10!white, draw = Plum!10!white, rounded corners = 0.2cm] (-2,0.6) rectangle (5.6,-2.6);
\draw[dotted,thick] (0,0) -- (5,0);
\draw[dotted,thick,red] (0,-1) -- (5,-1);
\draw[very thick] (5,0) -- (2.5,0);
\filldraw[black] (2.5,0) circle (2pt);
\draw[very thick,red] (0,-1) -- (2,-1);
\filldraw[red] (2,-1) circle (4pt);
\draw[dotted,thick,blue] (0,-2) -- (5,-2);
\draw[very thick,blue] (0,-2) -- (2,-2);
\draw[very thick,blue] (2.5,-2) -- (5,-2);
\filldraw[blue] (2,-2) circle (2pt);
\filldraw[blue] (2.5,-2) circle (2pt);
\node[right] at (-1.9,-2) {$R\ominus\tilde I$};
\node[right] at (-1.5,0) {$R$};
\node[right] at (-1.5,-1) {$\tilde I$};
\begin{scope}[xshift=10cm]
\filldraw[fill = Plum!10!white, draw = Plum!10!white, rounded corners = 0.2cm] (-2,0.6) rectangle (5.6,-2.6);
\draw[dotted,thick] (0,0) -- (5,0);
\draw[dotted,thick,red] (0,-1) -- (5,-1);
\draw[very thick] (5,0) -- (2.5,0);
\filldraw[black] (2.5,0) circle (2pt);
\draw[very thick,red] (0,-1) -- (3,-1);
\filldraw[red] (3,-1) circle (4pt);
\draw[dotted,thick,blue] (0,-2) -- (5,-2);
\draw[very thick,blue] (0,-2) -- (2.5,-2);
\draw[very thick,blue] (3,-2) -- (5,-2);
\filldraw[blue] (3,-2) circle (2pt);
\filldraw[blue] (2.5,-2) circle (2pt);
\node[right] at (-1.9,-2) {$R\ominus\tilde I$};
\node[right] at (-1.9,0) {$R$};
\node[right] at (-1.9,-1) {$\tilde I$};
\end{scope}
\end{scope}
\end{tikzpicture}
\end{center}}
\noindent In these cases, as $\tilde u_a$ increases, we have
\EQ{
\frac{\partial S_\text{gen.}}{\partial\tilde u_a}=\begin{cases} -\dot{\EuScript S}_a & \tilde u_a<u_i\ ,\\ 3\dot{\EuScript S}_a & \tilde u_a>u_i\ .\end{cases}
}
So in these cases, the entropy has a maximum at $\tilde u_a=u_i$. In the more detailed analysis of appendix \ref{a1}, we find that the generalized entropy behaves smoothly as $\tilde u_a$ moves through $u_i$ rather than the discontinuous na\"\i ve behaviour found here.

So a class of extrema exist where each element of $\partial\tilde I$ is mapped to a unique element of $\partial R$. As we have seen, only a subset of these extrema will actually be a minimum of the generalized entropy and so have a chance at dominating the entropy of $S(R)$. The above arguments imply that there are extrema of the generalized entropy when $\tilde u_a\approx u_i$. In appendix \ref{a1} we prove this in detail and show the next-to-leading corrections in the adiabatic limit are
\EQ{
u_a&=u_i-\frac1{2\pi T(u_i)}\log\lambda_a\ ,\\[5pt] v_a&=u_i-\frac1{2\pi T(u_i)}\log\frac{48\lambda_a(\SBH (u_i)-S_*)}{\cal N}\ .
\label{ono}
}
where $\lambda_a=\frac13,3$ for the for minimum/maximum cases above. 
\begin{figure}[ht]
\begin{center}
\begin{tikzpicture} [scale=1]
\filldraw[SkyBlue!20] (5.5,0.5) -- (7.5,2.5) -- (5.5,4.5) -- (1.5,0.5) -- cycle;  
\filldraw[SkyBlue!20] (0.5,4.5) -- (0.5,1.5) -- (4,5) --  (1,5) -- cycle;  
%
%
\draw[very thick, dotted] (0.5,0.5) -- (5,5);
\draw[black!20,very thick, dash dot] (5,5) -- (5,0.5); 
\filldraw[red] (3,4) circle (2.5pt);
\draw[thick,dotted,red] (5.5,4.5) -- (5,4) -- (5,2) -- (3,4);
%
%
%
\filldraw[black] (5.5,4.5) circle (2pt);
\draw[black,very thick] (5.5,4.5) -- (7.5,2.5);
\end{tikzpicture}
\caption{\footnotesize The relationship between a point in $\partial R$ in black and a QES $\partial I$ in red. At leading order in the adiabatic limit, the $U$ coordinate of the QES is determined by reflection $U\to-U$. The $v$ coordinate of the QES is equal to the $u$ coordinate of the point in $\partial R$ minus the scrambling time. This is subleading in the adiabatic limit but is nevertheless significant because it means that the QES lies close to the horizon.}
\label{fig4} 
\end{center}
\end{figure}
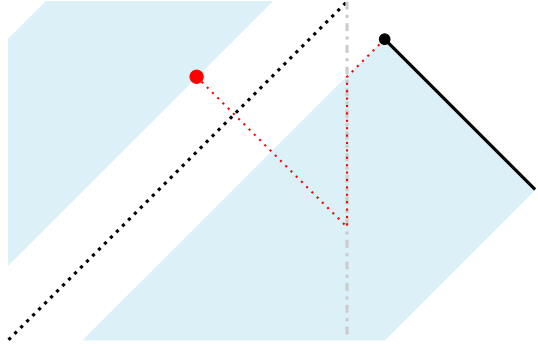

The difference 
\EQ{
\Delta t_\text{s}=u_i-v_a=\frac1{2\pi T(u_i)}\log\frac{48\lambda_a(\SBH (u_i)-S_*)}{\cal N}
}
is identified with the scrambling time of the black hole which is a subleading effect in the adiabatic limit: see figure \ref{fig4}. However, this subleading effect is responsible for ensuring that the QES are close to the horizon.

\section{Quantum information of the radiation}\label{s4}

In this section, we describe some applications of our `islands in the stream' formula \eqref{ger} for the entropy of any subset of the Hawking radiation as well as showing  that some of the fundamental constraints of quantum information theory are satisfied. We will suppose that the black hole forms from the collapse of a shockwave at $t=0$ and evaporates back to the extremal black hole at $t_\text{evap.}$. This is described in the JT gravity setting in \cite{Hollowood:2020cou}.

\subsection{The Page curve}

To derive the Page curve, we choose the interval $R=[0,u]$ to capture all the Hawking radiation emitted up to time $u$. We will take the extremal entropy $S_*$ to be negligible or vanishing, as in the Schwarzschild case. There are two possible saddles. The Hawking saddle has no island $I=\emptyset$ and entropy
\EQ{
S_\emptyset(R)=S_\text{rad}(R)=2(\ZZ_0-\ZZ_u)\ .
}
The second saddle has an island $I=\tilde R$, or $\tilde I=R$, with a QES just before the black hole is formed by an in-going shockwave or the origin for the Schwarzschild case, and so $R\ominus\tilde I=\emptyset$:
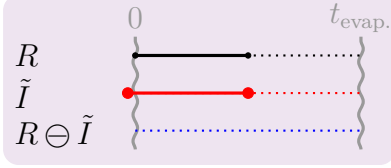
\begin{center}
\begin{tikzpicture}[scale=0.5]
\filldraw[fill = Plum!10!white, draw = Plum!10!white, rounded corners = 0.2cm] (-3.5,1.6) rectangle (6.9,-2.9);
\draw[decorate,very thick,black!40,decoration={snake,amplitude=0.03cm}] (6,-2.5) -- (6,0.5);
\draw[decorate,very thick,black!40,decoration={snake,amplitude=0.03cm}] (0,-2.5) -- (0,0.5);
\draw[dotted,thick] (0,0) -- (6,0);
\draw[dotted,thick,red] (0,-1) -- (6,-1);
\draw[dotted,thick,blue] (0,-2) -- (6,-2);
\draw[very thick] (3,0) -- (0,0);
\filldraw[black] (3,0) circle (2pt);
\filldraw[black] (0,0) circle (2pt);
\draw[very thick,red] (3,-1) -- (-0.2,-1);
\filldraw[red] (3,-1) circle (4pt);
\filldraw[red] (-0.2,-1) circle (4pt);
\node[right] at (-3.5,0) {$R$};
\node[right] at (-3.5,-1) {$\tilde I$};
\node[right] at (-3.5,-2) {$R\ominus\tilde I$};
\node[black!40] at (0,1) {$0$};
\node[black!40] at (6,1) {$t_\text{evap.}$};
\end{tikzpicture}
\end{center}
\noindent Hence, the only contribution to the entropy comes from the QES in the interval $[0,t_\text{evap.}]$
\EQ{
S_I(R)=\ZZ_u\ .
}

The entropy of $R$ is then a competition
\EQ{
S(R)=\min\big(2(\ZZ_0-\ZZ_u),\ZZ_u\big)\ .
}
For early times the Hawking saddle dominates but at the Page time
\EQ{
\ZZ_{u_\text{Page}}=\frac23\ZZ_0\ ,
}
there is a transition to the island saddle. Since we are assuming $S_*\approx0$, the final entropy at $u=t_\text{evap.}$ vanishes as expected on the basis of unitarity.\footnote{In the extremal case, or in JT gravity, with non-vanishing $S_*$, the details are more intricate. We can choose to collect the radiation in $R=[0,u]$ or the semi-infinite interval $R=[-\infty,u]$. The results are not the same even though no radiation is emitted before $u=0$. In the former case, one finds $S([0,u])=\min(2(\ZZ_0-\ZZ_u),S_*+\ZZ_u)$, where the island saddle has $\tilde I=[0^-,u]$. In the latter case, $S([-\infty,u])=\min(S_*+2(\ZZ_0-\ZZ_u),\ZZ_u)$ and in this case the Hawking saddle has an island $\tilde I=[-\infty,0^-]$ which lies outside the region $[0,t_\text{evap.}]$ and the island saddle has a semi-infinite island $\tilde I=[-\infty,u]$. These two scenarios have a different Page time.}
 
\subsection{A single interval}

Consider a single interval $R=[u_1,u_2]$, again assuming $S_*\approx0$. It has a Hawking saddle with entropy
\EQ{
S_\emptyset(R)=2(\ZZ_1-\ZZ_2)\ ,
}
and two possible island saddles:
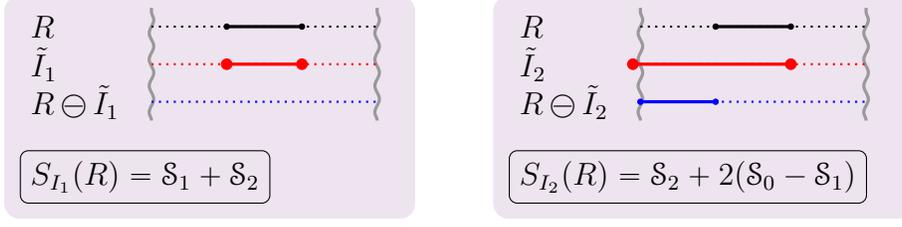
\begin{center}
\begin{tikzpicture} [scale=0.5]
\filldraw[fill = Plum!10!white, draw = Plum!10!white, rounded corners = 0.2cm] (-3.9,0.8) rectangle (7,-5.1);
\draw[decorate,very thick,black!40,decoration={snake,amplitude=0.03cm}] (6,-2.5) -- (6,0.5);
\draw[decorate,very thick,black!40,decoration={snake,amplitude=0.03cm}] (0,-2.5) -- (-0,0.5);
\draw[dotted,thick] (0,0) -- (6,0);
\draw[dotted,thick,red] (0,-1) -- (6,-1);
\draw[dotted,thick,blue] (0,-2) -- (6,-2);
\draw[very thick] (2,0) -- (4,0);
\filldraw[black] (2,0) circle (2pt);
\filldraw[black] (4,0) circle (2pt);
\draw[very thick,red] (2,-1) -- (4,-1);
\filldraw[red] (2,-1) circle (4pt);
\filldraw[red] (4,-1) circle (4pt);
\node[right] at (-3.5,0) {$R$};
\node[right] at (-3.5,-1) {$\tilde I_1$};
\node[right] at (-3.5,-2) {$R\ominus\tilde I_1$};
\node[right,draw=black,rounded corners=3pt] at (-3.5,-4) {$S_{I_1}(R)=\ZZ_1+\ZZ_2$};
\begin{scope}[xshift=13cm]
\filldraw[fill = Plum!10!white, draw = Plum!10!white, rounded corners = 0.2cm] (-3.9,0.8) rectangle (7.2,-5.1);
\draw[decorate,very thick,black!40,decoration={snake,amplitude=0.03cm}] (6,-2.5) -- (6,0.5);
\draw[decorate,very thick,black!40,decoration={snake,amplitude=0.03cm}] (0,-2.5) -- (-0,0.5);
\draw[dotted,thick] (0,0) -- (6,0);
\draw[dotted,thick,red] (0,-1) -- (6,-1);
\draw[dotted,thick,blue] (0,-2) -- (6,-2);
\draw[very thick] (2,0) -- (4,0);
\filldraw[black] (2,0) circle (2pt);
\filldraw[black] (4,0) circle (2pt);
\draw[very thick,red] (4,-1) -- (-0.2,-1);
\filldraw[red] (4,-1) circle (4pt);
\filldraw[red] (-0.2,-1) circle (4pt);
\draw[very thick,blue] (0,-2) -- (2,-2);
\filldraw[blue] (0,-2) circle (2pt);
\filldraw[blue] (2,-2) circle (2pt);
\node[right] at (-3.5,0) {$R$};
\node[right] at (-3.5,-1) {$\tilde I_2$};
\node[right] at (-3.5,-2) {$R\ominus\tilde I_2$};
\node[right,draw=black,rounded corners=3pt] at (-3.5,-4) {$S_{I_2}(R)=\ZZ_2+2(\ZZ_0-\ZZ_1)$};
\end{scope}
\end{tikzpicture}
\end{center}
\noindent The island $I_2$ starts on a QES just before the formation of the black hole. This is the QES of the extremal black hole (or the origin of the polar coordinates in the Schwarzschild case) but we are assuming that $S_*\approx0$. Hence, there is a competition between 3 saddles:
\EQ{
S(R)=\min\big(2(\ZZ_1-\ZZ_2),\ZZ_1+\ZZ_2,\ZZ_2+2(\ZZ_0-\ZZ_1)\big)\ .
}
The final saddle can only dominate if the point $u_1$ is sufficiently close to 0, that is
\EQ{
\ZZ_1>\frac23\ZZ_0\ .
\label{yee}
}
If $\ZZ_1$ does not satisfy this inequality then the Hawking saddle dominates when the interval is sufficiently small, specifically
\EQ{
\Delta_R\equiv 3\ZZ_2-\ZZ_1>0\ .
\label{saq}
}

\subsection{Unitarity}\label{s4.3}

Consider the single interval in the last section and define the complementary region $A$ which consists of two separate intervals:
\EQ{
A=[0,u_1]\cup[u_2,t_\text{evap.}]\ .
}
This complementary region has various saddles. First of all, the no-island saddle,
\EQ{
S_\emptyset(A)=2(\ZZ_0-\ZZ_1+\ZZ_2)\ .
}
The are 4 possible islands saddles:
\begin{center}
\begin{tikzpicture} [scale=0.5]
\filldraw[fill = Plum!10!white, draw = Plum!10!white, rounded corners = 0.2cm] (-3.9,0.8) rectangle (7,-5.1);
\draw[decorate,very thick,black!40,decoration={snake,amplitude=0.03cm}] (6,-2.5) -- (6,0.5);
\draw[decorate,very thick,black!40,decoration={snake,amplitude=0.03cm}] (0,-2.5) -- (0,0.5);
\draw[dotted,thick] (0,0) -- (6,0);
\draw[dotted,thick,red] (0,-1) -- (6,-1);
\draw[dotted,thick,blue] (0,-2) -- (6,-2);
\draw[very thick] (2,0) -- (0,0);
\draw[very thick] (4,0) -- (6,0);
\filldraw[black] (0,0) circle (2pt);
\filldraw[black] (2,0) circle (2pt);
\filldraw[black] (4,0) circle (2pt);
\filldraw[black] (6,0) circle (2pt);
\draw[very thick,red] (4,-1) -- (6,-1);
\filldraw[red] (4,-1) circle (4pt);
\filldraw[red] (6,-1) circle (4pt);
\draw[very thick,blue] (2,-2) -- (0,-2);
\filldraw[blue] (2,-2) circle (2pt);
\filldraw[blue] (0,-2) circle (2pt);
\node[right] at (-3.5,0) {$A$};
\node[right] at (-3.5,-1) {$\tilde I'_1$};
\node[right] at (-3.5,-2) {$A\ominus\tilde I'_1$};
\node[right,draw=black,rounded corners=3pt] at (-3.5,-4) {$S_{I'_1}(A)=\ZZ_2+2(\ZZ_0-\ZZ_1)$};
\begin{scope}[xshift=15cm]
\filldraw[fill = Plum!10!white, draw = Plum!10!white, rounded corners = 0.2cm] (-3.9,0.8) rectangle (7,-5.1);
\draw[decorate,very thick,black!40,decoration={snake,amplitude=0.03cm}] (6,-2.5) -- (6,0.5);
\draw[decorate,very thick,black!40,decoration={snake,amplitude=0.03cm}] (0,-2.5) -- (0,0.5);
\draw[dotted,thick] (0,0) -- (6,0);
\draw[dotted,thick,red] (0,-1) -- (6,-1);
\draw[dotted,thick,blue] (0,-2) -- (6,-2);
\draw[very thick] (2,0) -- (0,0);
\draw[very thick] (2,0) -- (0,0);
\draw[very thick] (4,0) -- (6,0);
\filldraw[black] (0,0) circle (2pt);
\filldraw[black] (2,0) circle (2pt);
\filldraw[black] (4,0) circle (2pt);
\filldraw[black] (6,0) circle (2pt);
\draw[very thick,red] (2,-1) -- (-0.2,-1);
\filldraw[red] (-0.2,-1) circle (4pt);
\filldraw[red] (2,-1) circle (4pt);
\draw[very thick,blue] (4,-2) -- (6,-2);
\filldraw[blue] (4,-2) circle (2pt);
\filldraw[blue] (6,-2) circle (2pt);
\node[right] at (-3.5,0) {$A$};
\node[right] at (-3.5,-1) {$\tilde I'_2$};
\node[right] at (-3.5,-2) {$A\ominus\tilde I'_2$};
\node[right,draw=black,rounded corners=3pt] at (-3.5,-4) {$S_{I'_2}(A)=\ZZ_1+2\ZZ_2$};
\end{scope}
\begin{scope}[xshift=0cm,yshift=-7cm]
\filldraw[fill = Plum!10!white, draw = Plum!10!white, rounded corners = 0.2cm] (-3.9,0.8) rectangle (7,-5.1);
\draw[decorate,very thick,black!40,decoration={snake,amplitude=0.03cm}] (6,-2.5) -- (6,0.5);
\draw[decorate,very thick,black!40,decoration={snake,amplitude=0.03cm}] (0,-2.5) -- (0,0.5);
\draw[dotted,thick] (0,0) -- (6,0);
\draw[dotted,thick,red] (0,-1) -- (6,-1);
\draw[dotted,thick,blue] (0,-2) -- (6,-2);
\draw[very thick] (2,0) -- (0,0);
\draw[very thick] (2,0) -- (0,0);
\draw[very thick] (4,0) -- (6,0);
\filldraw[black] (0,0) circle (2pt);
\filldraw[black] (2,0) circle (2pt);
\filldraw[black] (4,0) circle (2pt);
\filldraw[black] (6,0) circle (2pt);
\draw[very thick,red] (2,-1) -- (-0.2,-1);
\filldraw[red] (-0.2,-1) circle (4pt);
\filldraw[red] (2,-1) circle (4pt);
\draw[very thick,red] (4,-1) -- (6,-1);
\filldraw[red] (4,-1) circle (4pt);
\filldraw[red] (6,-1) circle (4pt);
\node[right] at (-3.5,0) {$A$};
\node[right] at (-3.5,-1) {$\tilde I'_3$};
\node[right] at (-3.5,-2) {$A\ominus\tilde I'_3$};
\node[right,draw=black,rounded corners=3pt] at (-3.5,-4) {$S_{I'_3}(A)=\ZZ_1+\ZZ_2$};
\end{scope}
\begin{scope}[xshift=15cm,yshift=-7cm]
\filldraw[fill = Plum!10!white, draw = Plum!10!white, rounded corners = 0.2cm] (-3.9,0.8) rectangle (7,-5.1);
\draw[decorate,very thick,black!40,decoration={snake,amplitude=0.03cm}] (6,-2.5) -- (6,0.5);
\draw[decorate,very thick,black!40,decoration={snake,amplitude=0.03cm}] (0,-2.5) -- (0,0.5);
\draw[dotted,thick] (0,0) -- (6,0);
\draw[dotted,thick,red] (0,-1) -- (6,-1);
\draw[dotted,thick,blue] (0,-2) -- (6,-2);
\draw[very thick] (2,0) -- (0,0);
\draw[very thick] (2,0) -- (0,0);
\draw[very thick] (4,0) -- (6,0);
\filldraw[black] (0,0) circle (2pt);
\filldraw[black] (2,0) circle (2pt);
\filldraw[black] (4,0) circle (2pt);
\filldraw[black] (6,0) circle (2pt);
\draw[very thick,red] (6,-1) -- (-0.2,-1);
\filldraw[red] (-0.2,-1) circle (4pt);
\filldraw[red] (6,-1) circle (4pt);
\draw[very thick,blue] (2,-2) -- (4,-2);
\filldraw[blue] (2,-2) circle (2pt);
\filldraw[blue] (4,-2) circle (2pt);
\node[right] at (-3.5,0) {$A$};
\node[right] at (-3.5,-1) {$\tilde I'_4$};
\node[right] at (-3.5,-2) {$A\ominus\tilde I'_4$};
\node[right,draw=black,rounded corners=3pt] at (-3.5,-4) {$S_{I'_4}(A)=2(\ZZ_1-\ZZ_2)$};
\end{scope}
\end{tikzpicture}
\end{center}

Since we are assuming the $S_*\approx0$, the state of the radiation is a pure state, or approximately so, and the no-island saddle can never dominate because $S_\emptyset(A)>S_{I'_1}(A)$. In addition, the $I'_2$ saddle can never dominate because $S_{I'_2}(A)>S_{I'_3}(A)$. 

When the state of the  radiation is pure, it must be that $S(A)=S(R)$. Indeed, we find perfect matching of the saddles that can dominate,
\EQ{
S_\emptyset(R)=S_{I'_4}(A)\  ,\qquad S_{I_1}(R)=S_{I'_3}(A)\ ,\qquad S_{I_2}(R)=S_{I'_1}(A)\ ,
}
providing a highly non-trivial test of unitarity. What is interesting is that for all these saddles the reflections of the islands are complementary, meaning, e.g.~$\tilde I_1\cap\tilde I'_3=\emptyset$ and $\tilde  I_1\cup \tilde I'_3=[0,t_\text{evap.}]$.
 
\subsection{Mutual information of two intervals}
\label{sub:mutual_inf}

Consider two intervals $R_1$  and $R_2$. It  is clear  that the mutual information $I(R_1,R_2)$ will only be non-vanishing in a saddle with an island that straddles both $R_1$  and $R_2$:\footnote{Note that the  mutual information is UV safe quantity because the cut off terms cancel out.}
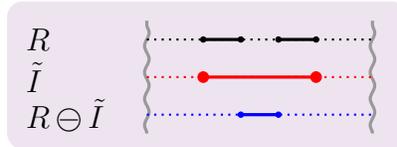
\begin{center}
	\begin{tikzpicture} [scale=0.5]
\filldraw[fill = Plum!10!white, draw = Plum!10!white, rounded corners = 0.2cm] (-3.7,1) rectangle (6.9,-2.9);
	\draw[decorate,very thick,black!40,decoration={snake,amplitude=0.03cm}] (6,-2.5) -- (6,0.5);
	\draw[decorate,very thick,black!40,decoration={snake,amplitude=0.03cm}] (0,-2.5) -- (0,0.5);
	\draw[dotted,thick] (0,0) -- (6,0);
	\draw[dotted,thick,red] (0,-1) -- (6,-1);
	\draw[dotted,thick,blue] (0,-2) -- (6,-2);
	\draw[very thick] (1.5,0) -- (2.5,0);
	\draw[very thick] (3.5,0) -- (4.5,0);
	\filldraw[black] (1.5,0) circle (2pt);
	\filldraw[black] (2.5,0) circle (2pt);
	\filldraw[black] (3.5,0) circle (2pt);
	\filldraw[black] (4.5,0) circle (2pt);
	\draw[very thick,red] (1.5,-1) -- (4.5,-1);
	\filldraw[red] (1.5,-1) circle (4pt);
	\filldraw[red] (4.5,-1) circle (4pt);
	\draw[very thick,blue] (2.5,-2) -- (3.5,-2);
	\filldraw[blue] (2.5,-2) circle (2pt);
	\filldraw[blue] (3.5,-2) circle (2pt);
	%
	%
	%
	\node[right] at (-3.5,0) {$R$};
	\node[right] at (-3.5,-1) {$\tilde I$};
	\node[right] at (-3.5,-2) {$R\ominus\tilde I$};
	\end{tikzpicture}
\end{center}
In order to calculate the mutual information, we need to know whether $R_1$ or $R_2$ are, themselves, in their island saddle $\tilde I_1=R_1$ or $\tilde I_2=R_2$, respectively. For simplicity, let us assume that $R_1$  is not too close to 0, more specifically the condition  \eqref{yee} is satisfied, although the result is not altered if we relax this condition. Assuming this is true, the condition for both intervals to be in their Hawking saddles is determined by the condition \eqref{saq} on $\Delta_{R_1}$ and $\Delta_{R_2}$. There are 4 possible cases (denoting the interval in between $R_1$ and $R_2$  as $P$):
\begin{enumerate}
\item $\tilde I_1=\emptyset$ and $\tilde I_2=\emptyset$, 
\EQ{
	I(R_1,R_2)=\max(0,\Delta_P-\Delta_{R_1}-\Delta_{R_2})\ .
} 
\item $\tilde I_1=\emptyset$ and $\tilde I_2=R_2$, 
\EQ{
	I(R_1,R_2)=\max(0,\Delta_P-\Delta_{R_1})\ .
} 
\item $\tilde I_1=R_1$ and $\tilde I_2=\emptyset$,  
\EQ{
	I(R_1,R_2)=\max(0,\Delta_P-\Delta_{R_2})\ .
} 
\item $\tilde I_1=R_1$ and $\tilde I_2=R_2$, 
\EQ{
	I(R_1,R_2)=\max(0,\Delta_P)\ .
}
\end{enumerate}
It is remarkable that we can amalgamate all these different cases into a single expression for the mutual information which automatically takes care of what saddles dominate for $R_1$ and $R_2$ separately,
\EQ{
	I(R_1,R_2)=\max(0,\Delta_P-\max(\Delta_{R_1}+\Delta_{R_2},\Delta_{R_1},\Delta_{R_2},0))\ .
 \label{eq:MI_result}}
Notice that the mutual information is manifestly positive and so subadditivity is satisfied.

\subsection{Araki-Lieb inequality}
\label{sub:araki-lieb}

The Araki-Lieb or triangle inequality $S(R_1 \cup R_2) \ge |S(R_1)-S(R_2)|$ \cite{Araki:1970ba} can be simply rephrased as an upper bound on the mutual information
\begin{equation}
2 \min (S(R_1) , S(R_2)) \ge I(R_1,R_2) \, .
\end{equation}
We will prove this inequality using the same setup of the previous section.\footnote{In particular, we will assume that $R_1$ is not too early in the sense of the condition \eqref{yee}. The condition can be relaxed, complicating the analysis without changing the conclusion.} Since to compute the mutual information of \eqref{eq:MI_result} we have to take the minimum\footnote{More precisely, minus the maximum.} among several possibilities, and assuming that $I(R_1,R_2)\neq0$, it is enough to prove that $2 S(R_{1,2})$ is bigger than one of these quantities to prove that the Araki-Lieb inequality is satisfied.

Let us begin by considering $R_1$. We always have that the entropy of an interval is bigger than half of its no-island saddle: $2S(R_1)\geq S_\emptyset (R_1)$. Using this fact, we have the following chain of inequalities:
\begin{equation}
2S(R_1) \ge S_\emptyset (R_1) \ge \Delta_P-\Delta_{R_1} \ge I(R_1,R_2) \, .
\end{equation}
For $R_2$ it is more convenient to consider the saddles separately. In particular, we have
\begin{equation}
2S_\emptyset(R_2) \ge \Delta_P - \Delta_{R_2} \, , \qquad 2S_I(R_2) \ge \Delta_P \, \quad \Rightarrow \quad 2S(R_2)  \ge I(R_1,R_2) \, .
\end{equation}
This proves that the Araki-Lieb inequality is satisfied. \color{black}

\subsection{Monogamy of mutual information}
\label{sub:CMI}

The mutual information of two intervals $R_1,R_2$, given a third interval $R_3$,  is quantified by the conditional mutual information: 
\begin{equation}
\label{eq:R_3MI_def}
I(R_1,R_2|R_3)= S(R_1 \cup R_3)+S(R_2 \cup R_3)-S(R_3)-S(R_1\cup R_2 \cup R_3) \, .
\end{equation}
The constraint of strong subadditivity is the condition that the conditional mutual information is positive. We will prove a more stringent inequality, also known as monogamy of mutual information\footnote{The proof of strong subadditivity property of the generalised entropy was given in \cite{Akers:2019lzs}. In the same paper it is also argued that the generalised entropy satisfies monogamy of mutual information if the von Neumann entropy of quantum fields does. This is the case for the entropy of thermal radiation \eqref{dup}.}:
\begin{equation}
\label{eq:MI<R_3MI_def}
I(R_1,R_2|R_3)\ge I(R_1,R_2)\ .
\end{equation} 
If such condition is verified it would suggest that the correlations among the Hawking radiation are mostly quantum. Indeed the minimal $I(R_1,R_2|R_3)$ among all the possible choices of $R_3 \neq R_{1,2}$ is a measure of the entanglement between $R_1$ and $R_2$ \cite{CW}. On the other hand, the mutual information is believed to be an upper bound for both classical and quantum correlations, and therefore \eqref{eq:MI<R_3MI_def} is telling us that there is no room for classical correlations.

In order to prove \eqref{eq:MI<R_3MI_def}, we can write it as a condition that is completely symmetric in $R_1$, $R_2$ and $R_3$:
\begin{equation}
\label{eq:MI<CMI}
S(R_1)+S(R_2)+S(R_3)+S(R_1\cup R_2 \cup R_3) \le S(R_1 \cup R_2)+S(R_1 \cup R_3)+S(R_2 \cup R_3) \, .
\end{equation}
We will assume that the first interval $R_1$ is subject to the constraint \eqref{yee}. The condition can be relaxed at the result of a more complicated analysis with the same conclusion.

We will now divide the proof into cases depending on how $S (R_1 \cup R_2 \cup R_3)$ factorizes as governed by which island dominates the entropy. There are four possible cases:
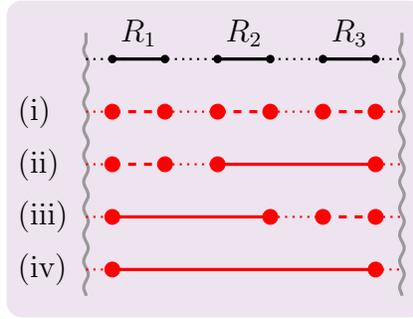
\begin{center}
	\begin{tikzpicture} [scale=0.7]
\filldraw[fill = Plum!10!white, draw = Plum!10!white, rounded corners = 0.2cm] (-1.5,1.1) rectangle (6.4,-4.9);
	\draw[decorate,very thick,black!40,decoration={snake,amplitude=0.03cm}] (6,-4.5) -- (6,0.5);
	\draw[decorate,very thick,black!40,decoration={snake,amplitude=0.03cm}] (0,-4.5) -- (0,0.5);
	\draw[dotted,thick] (0,0) -- (6,0);
	\draw[dotted,thick,red] (0,-1) -- (0.5,-1);
	\draw[dotted,thick,red] (1.5,-1) -- (2.5,-1);
	\draw[dotted,thick,red] (3.5,-1) -- (4.5,-1);
	\draw[dotted,thick,red] (5.5,-1) -- (6,-1);	
	\draw[dotted,thick,red] (0,-2) -- (0.5,-2);
	\draw[dotted,thick,red] (1.5,-2) -- (2.5,-2);
	\draw[dotted,thick,red] (5.5,-2) -- (6,-2);
	\draw[dotted,thick,red] (0,-3) -- (0.5,-3);
	\draw[dotted,thick,red] (3.5,-3) -- (4.5,-3);
	\draw[dotted,thick,red] (5.5,-3) -- (6,-3);
	\draw[dotted,thick,red] (0,-4) -- (6,-4);
	\draw[very thick] (2.5,0) -- (3.5,0);
	\draw[very thick] (0.5,0) -- (1.5,0);
	\draw[very thick] (4.5,0) -- (5.5,0);
	\filldraw[black] (0.5,0) circle (2pt);
	\filldraw[black] (1.5,0) circle (2pt);
	\filldraw[black] (2.5,0) circle (2pt);
	\filldraw[black] (3.5,0) circle (2pt);
	\filldraw[black] (4.5,0) circle (2pt);
	\filldraw[black] (5.5,0) circle (2pt);
	\filldraw[red] (0.5,-1) circle (4pt);
	\filldraw[red] (1.5,-1) circle (4pt);
	\filldraw[red] (2.5,-1) circle (4pt);
	\filldraw[red] (3.5,-1) circle (4pt);
	\filldraw[red] (4.5,-1) circle (4pt);
	\filldraw[red] (5.5,-1) circle (4pt);
	\draw[very thick,red, dashed] (2.5,-1) -- (3.5,-1);
	\draw[very thick,red,dashed] (0.5,-1) -- (1.5,-1);
	\draw[very thick,red,dashed] (4.5,-1) -- (5.5,-1);
	\filldraw[red] (0.5,-2) circle (4pt);
	\filldraw[red] (1.5,-2) circle (4pt);
	\filldraw[red] (2.5,-2) circle (4pt);
	\filldraw[red] (5.5,-2) circle (4pt);
	\draw[very thick,red,dashed] (0.5,-2) -- (1.5,-2);
	\draw[very thick,red] (2.5,-2) -- (5.5,-2);
	\filldraw[red] (0.5,-3) circle (4pt);
	\filldraw[red] (3.5,-3) circle (4pt);
	\filldraw[red] (4.5,-3) circle (4pt);
	\filldraw[red] (5.5,-3) circle (4pt);
	\draw[very thick,red] (0.5,-3) -- (3.5,-3);
	\draw[very thick,red,dashed] (4.5,-3) -- (5.5,-3);
	\draw[very thick,red] (0.5,-4) -- (5.5,-4);
	\filldraw[red] (0.5,-4) circle (4pt);
	\filldraw[red] (5.5,-4) circle (4pt);
	\node at (1,0.5) {$R_1$};
	\node at (3,0.5) {$R_2$};
	\node at (5,0.5) {$R_3$};
	\node[right] at (-1.5,-1) {(i)};
	\node[right] at (-1.5,-2) {(ii)};
	\node[right] at (-1.5,-3) {(iii)};
	\node[right] at (-1.5,-4) {(iv)};
	\end{tikzpicture}
\end{center}
where the dashed segments can admit either a $\emptyset$ or $\tilde R_j$ saddle.

\noindent (i) In this case, it is clear that $S_\text{(i)}= S(R_1) + S(R_2) + S(R_3)$ and $S(R_i\cup R_j)=S(R_i)+S(R_j)$ and therefore \eqref{eq:MI<CMI} is trivially satisfied as an equality.

\noindent (ii) In this case $S_\text{(ii)}= S(R_1) + S(R_2 \cup R_3)$ and if this island dominates then it follows that $S(R_1 \cup R_2) = S(R_1) + S(R_2)$ and $S(R_1 \cup R_3) = S(R_1) + S(R_3)$, in which case  \eqref{eq:MI<CMI} is satisfied as an equality. 

\noindent (iii) This case is argued in an identical way to (ii).

\noindent (iv) This is  the most challenging case because $S_\text{(iv)}$ does not factorize and we have to go through the sub-cases according to which of the $S(R_i\cup R_j)$ are in their island saddle. We denote the set of pairs that are in their island saddle as $\EuScript U$. Of the 8 possibilities, the case $\EuScript U=\{12,23\}$ cannot occur because those two pairs would imply that $S(R_1\cup R_3)$ was also in its island saddle. Pairs that are not in $\EuScript U$ satisfy $S(R_i\cup R_j)=S(R_i)+S(R_j)$. Taking the seven remaining possibilities seriatim:
\begin{enumerate}[label=\protect\circled{\arabic*}]
\item $\EuScript U=\emptyset$. In this case \eqref{eq:MI<CMI} reads
\EQ{
S_\text{(iv)}\overset?\leq S(R_1)+S(R_2)+S(R_3)\equiv S_\text{(i)}\ ,
}
which is satisfied because we are assuming that $S_\text{(iv)}<S_\text{(i)}$.
\item $\EuScript U=\{12\}$. In this case \eqref{eq:MI<CMI} reads
\EQ{
S_\text{(iv)}\overset?\leq S(R_1\cup R_2)+S(R_3)\equiv S_\text{(iii)}\ ,
}
which is satisfied because we are assuming that $S_\text{(iv)}<S_\text{(iii)}$.
\item $\EuScript U=\{23\}$. This case is proved in an identical way to case 2.
\item $\EuScript U=\{13\}$. In this case \eqref{eq:MI<CMI} reads
\EQ{
S_\text{(iv)}\overset?\leq S(R_2)+S(R_1\cup R_3)\ .
\label{kiw}
}
But a useful identity is that when $S(R_1\cup R_3)$ is in its island saddle then
\EQ{
S(R_1\cup R_3)=S_\text{(iv)}+S_\emptyset(R_2)
\label{vvp}
}
and so \eqref{kiw} beomes
\EQ{
0\overset?\leq S(R_2)+S_\emptyset(R_2)\ ,
\label{kiw2}
}
which is clearly satisfied.
\item $\EuScript U=\{12,13\}$. In this case, using \eqref{vvp}, \eqref{eq:MI<CMI} reads
\EQ{
S(R_1)-S_\emptyset(R_2)\overset?\leq S(R_1\cup R_2)\ .
\label{yrp}
}
The Araki-Lieb inequality and the fact that $S(R_2)\leq S_\emptyset(R_2)$ lead to
\EQ{
S(R_1\cup R_2)\geq S(R_1)-S(R_2)\geq S(R_1)-S_\emptyset(R_2)
}
which implies \eqref{yrp}.
\item $\EuScript U=\{13,23\}$. This case is proved in the same way as case 5 above.
\item $\EuScript U=\{12,13,23\}$. In this case, using  \eqref{vvp}, \eqref{eq:MI<CMI} reads
\EQ{
0&\overset?\leq \big(S(R_1\cup R_2)-S(R_1)+S(R_2)\big)\\ &+\big(S(R_2\cup R_3)-S(R_2)+S(R_3)\big)+\big(S_\emptyset(R_2)-S(R_2)\big)\ .
\label{kiw4}
}
The 3 terms in brackets are positive, the first two as a consequence of the Araki-Lieb inequality.
\end{enumerate}

\section{Discussion}\label{s5}

We have shown how the fact that black holes evaporate very slowly, at least for most of their life, can be exploited to write a very simple recipe for computing the von Neumann entropy of any subset of the Hawking radiation  with intervals that are suitably large. The most striking feature of the result is that the entanglement effect of the island can be taken care of in terms of its mirror image in the outgoing radiation. This is the `island in the stream'. This yields a simple pictorial description (in a geometrical optics type limit)  of how correlations are encoded in the Hawking radiation in the bath whilst also providing a simple method of identifying multiple saddle points that can potentially contribute to the evolution of the entanglement strutcure of the radiation.
 
The formalism will be exploited in a companion paper to investigate the Hayden-Preskill scenario of throwing a diary into a black hole and asking when the information is returned in the radiation \cite{Hayden:2007cs} and the Harlow-Hayden process of distilling the purifier of a late portion of the Hawking radiation in the early radiation \cite{Harlow:2013tf}. Another feature that can be investigated is the existence of obstructions to decoding the state of the radiation and black hole known as a python's lunch \cite{Brown:2019rox}.

In this work we have avoided the complications of having a non-trivial greybody factor. The extension of the formalism to include a non-trivial greybody factor is addressed in \cite{Grey} where we show that there is a very simple generalization of the `islands in the stream' formula for the entropy \eqref{ger}. Essentially one replaces $S_\text{rad}(R\ominus\tilde I)$ with the thermodynamic entropy of $R\ominus\tilde I$ with a non-trivial greybody factor $\Gamma(\omega)\neq1$, i.e.~\eqref{hxx} with occupation number \eqref{kex}.

\vspace{0.5cm}
\begin{center}{\it Acknowledgments}\end{center}
\vspace{0.2cm}
TJH, AL and SPK acknowledge support from STFC grant ST/T000813/1. NT acknowledges the support of an STFC Studentship.

\appendix
\appendixpage

\section{Solving for the QES}
\label{a1}

In this appendix we implement the island prescription to the find a class of islands that extremize the generalized entropy for a given set of intervals $R$ of the Hawking radiation defined on $\mathscr I^+$ associated to null coordinates $\partial R=\{u_i\}$ with $u_1>u_2>\ldots>u_n$. We will not be able to claim that our set of extremal are complete but it is a reasonable hypothesis that our set contains that ones that dominate the entropy in the adiabatic limit. Indeed, the analysis in  appendix C of \cite{Brown:2019rox} finds an extremum that is not in the class considered here, however, this is a maximum of the generalized entropy.

Let us consider an island $I$ consisting of a number of QES $\partial I=\{(U_a,V_a),\ a=1,\ldots,p\}$. Note that $p$ and $n$ are even. The points $\partial  I$ and $\partial R$ are ordered so that
\EQ{
U_a\gg U_b\ ,\quad a>b\ ,\qquad |U_i|\ll  |U_j|\ ,\quad i<j\ ,
\label{bxx}
}
where $U_a>0$ (i.e.~inside the horizon) and $U_i<0$ (i.e.~outside the horizon). In addition,  
the intervals are sufficiently large that the thermodynamic limit for the entropy \eqref{ab2} applies.

We now turn to the generalized entropy in \eqref{guz2}. We will assume that the QES are close to the horizon in the sense that $U_aV_a\ll1$, a fact that we will prove ex post facto. The contribution from the QES can then be written using the universal near-horizon expression \eqref{nil},
The second term in \eqref{guz2} is the QFT entropy for free bosons in the Unruh vacuum,\footnote{This is proved  by exploiting  conformal invariance and the general results developed in \cite{Calabrese:2004eu}.}
\EQ{
S_\text{QFT}(R\cup I)&=-\frac{\cal N}6\sum_{a<b}(-1)^{a-b}\log\sigma_{ab}+\frac{\cal N}6\sum_{ai}(-1)^{a-i}\log\sigma_{ai}\\[5pt] &-\frac{\cal N}6\sum_{i<j}(-1)^{i-j}\log\sigma_{ij}-\frac{\cal N}{6}\sum_a\log\Omega_a-\frac{\cal N}6\sum_i\log\Omega_i\ .
\label{jew}
}
where $\sigma_{ab}=-(U_a-U_b)(v_a-v_b)$, etc. The last terms in \eqref{jew} are the conformal factors of the endpoints. For a point in $\partial R$, this arises from the conformal transformation $u\to U$:
\EQ{
\Omega_i^{-2}=\frac{\partial u_i}{\partial U_i}=\frac1{2\pi T(u_i)U_i}\ .
}
While for a point in $\partial I$, i.e.~a QES, there is the conformal factor of the metric \eqref{zik} and the conformal transformation $V\to v$:
\EQ{
\Omega_a^{-2}=\frac1{(1+U_aV_a)^2}\cdot\frac{\partial V_a}{\partial v_a}\approx 2\pi T(v_a)V_a\ ,
\label{pop2}
}
where we have assumed the near-horizon approximation $U_aV_a\ll1$.

Let us first extremize the generalized entropy with respect to $V_a$. It is apparent that the $\log(v_a-v_b)$ terms can be ignored because the contributions are subleading in the adiabatic limit. The $V_a$ derivative of $\SBH (v_a)$ in \eqref{nil} can be evaluated by using
\EQ{
\frac{d\SBH }{dV}=\frac1T\cdot\frac{dM}{dV}=-\frac{\cal N}{24V}\  ,
\label{nip}
}
where we used \eqref{her}. Hence, extremizing with respect to $V_a$ gives
\EQ{
2(\SBH (v_a)-S_*)U_a+\frac{\cal N}{24V_a}-\frac{\cal N}{12V_a}=0\ ,
\label{eq2}
}
where the second term comes from \eqref{nip} and the third from the conformal factor  \eqref{pop2}. This condition is precisely \eqref{hik2} so that $U_aV_a\ll1$ as anticipated in the adiabatic limit \eqref{adi}.

Now we extremize the generalized entropy with respect to $U_a$ yielding the coupled equations
\EQ{
2(\SBH (v_a)-S_*)V_a+\frac{\cal N}6\sum_{b(\neq a)}\frac{(-1)^{a-b}}{U_a-U_b}-
\frac{\cal N}6\sum_{j}\frac{(-1)^{a-j}}{U_a-U_j}=0\ .
\label{eq1}
}
In order to solve these equations we can exploit the fact that according to the condition \eqref{ab2} the $U$ coordinates of points in $\partial R$ are  well separated in the the sense that $|U_i/U_j|\ll 1$, for $i<j$. 

We proceed to make an ansatz that at leading order in the adiabatic approximation 
\EQ{
\tilde u_a=u_{\alpha(a)}+\cdots\ ,
\label{slm}
}
for a one-to-one map $\alpha$ of the QES to a subset of points in $\partial R$ that preserve the order, so $\alpha(a)>\alpha(b)$, for $a>b$. This ansatz means that a point in $\partial I$, i.e.~a QES, has a reflection under $U\to-U$, that is in the set $\partial R$, i.e.~$\partial\tilde I\subset\partial R$. So the islands in the stream have endpoints that lie in the set $\partial R$. 

Let  us write the next-to-leading order correction in the form 
\EQ{
U_a=-\lambda_aU_{\alpha(a)}\ ,
\label{fur2}
}
and solve for $\lambda_a$. The key insight is to appreciate the relative magnitudes of the $U$ coordinates of the QES and $\partial R$:
\EQ{
\begin{tikzpicture}[scale=1]
\node[right] at (0,0) {$|U_1|\ll \cdots \ll |U_{\alpha(a)}|\ll \cdots   \ll |U_{\alpha(b)}|\ll \cdots \ll |U_n|$};
\draw[thick,dotted,<->] (3,-0.4) -- (3,-1.2);
\draw[thick,dotted,<->] (5.5,-0.4) -- (5.5,-1.2);
\node[right] at (1.5,-1.5) {$\cdots \ll U_a\ll~ ~\cdots~~   \ll U_b\ll\cdots$};
\end{tikzpicture}
\label{ut}
}
for $a>b$, where the dotted arrows indicate terms of the same order.  Let us consider \eqref{eq1}, and use the relative magnitudes above. This leads to the tractable equation,
\EQ{
2(\SBH (v_a)-S_*)V_a=\frac{\cal N}6\begin{cases} (U_a-U_{\alpha(a)})^{-1} & a+\alpha(a)\in\text{even}\ ,\\[5pt]
U_a^{-1}-(U_a-U_{\alpha(a)})^{-1} & a+\alpha(a)\in\text{odd}\ .\end{cases}
}
Using \eqref{fur2}, along with \eqref{hik2}, it follows that
\EQ{
\lambda_a=\begin{cases} \frac13 & a+\alpha(a)\in\text{even}\ ,\\ 3 & a+\alpha(a)\in\text{odd}\ .
\end{cases}
}

So we have the solution for the QES at the next-to-leading order, precisely \eqref{ono} with $i$ identified with $\alpha(a)$. Notice at this order we can replace $v_a$ in $T(v_a)$ and $\SBH (v_a)$ with $u_{\alpha(a)}$.

With the leading order behaviour, we can compute the entropy of the island saddle. We separate this into four contributions:
\begin{enumerate}[label=\protect\circled{\arabic*}]
\item The contribution to $S_\text{QFT}(R\cup I)$ which is simply $S_\text{rad}(R)$ in the adiabatic limit.
\item The contribution to $S_\text{QFT}(R\cup I)$ from the island,
\EQ{
&-\frac{\cal N}6\sum_{a<b}(-1)^{a-b}\log(U_b-U_a)+\frac{\cal N}{12}\sum_a\log V_a\\
&=-\frac{\cal N}6\sum_{a<b}(-1)^{a-b}\log(U_{\alpha(a)}-U_{\alpha(b)})-\frac{\cal N}{12}\sum_a\log(-U_{\alpha(a)})
=S_\text{rad}(\tilde I)\ ,
}
to leading order, where $\tilde I$ is the reflection of the island under the mapping $U\to-U$, the `island in the stream'. Notice how the infalling sector provides a contribution that looks like a conformal factor because of the condition \eqref{hik2}, i.e.~$V_a\sim1/U_a$ at leading order.
\item Cross terms between the island and bath
\EQ{
\frac{\cal N}6\sum_{aj}(-1)^{a-j}\log(U_a-U_j)&=-\frac{\cal N}6\sum_{aj}(-1)^{a-j}\log|U_{\max(\alpha(a),j)}|\\ &=-2S_\text{rad}(R\cap\tilde I)\ .
}
\item Finally, there are the contributions from the QES $\sum_{\partial\tilde I}\SBH (u_{\partial\tilde I})$.
\end{enumerate}

The sum of the QFT contributions is
\EQ{
S_\text{rad}(R)+S_\text{rad}(\tilde I)-2S_\text{rad}(R\cap\tilde I)=S_\text{rad}(R\ominus\tilde I)
}
and so we have our result \eqref{ger}.


\begin{thebibliography}{99}
{\small

\bibitem{Penington:2019kki}
  G.~Penington, S.~H.~Shenker, D.~Stanford and Z.~Yang,
  ``Replica wormholes and the black hole interior,''
  \arXiv{1911.11977} [hep-th].
 
  
\bibitem{Almheiri:2019qdq}
A.~Almheiri, T.~Hartman, J.~Maldacena, E.~Shaghoulian and A.~Tajdini,
``Replica Wormholes and the Entropy of Hawking Radiation,''
JHEP \textbf{05} (2020), 013
[\arXiv{1911.12333} [hep-th]].

\bibitem{Hawking:1974sw}
S.~W.~Hawking,
``Particle Creation by Black Holes,''
Commun. Math. Phys. \textbf{43} (1975), 199-220
[erratum: Commun. Math. Phys. \textbf{46} (1976), 206]

\bibitem{Hawking:1976ra}
S.~W.~Hawking,
``Breakdown of Predictability in Gravitational Collapse,''
Phys. Rev. D \textbf{14} (1976), 2460-2473

\bibitem{Almheiri:2020cfm}
A.~Almheiri, T.~Hartman, J.~Maldacena, E.~Shaghoulian and A.~Tajdini,
``The entropy of Hawking radiation,''
[\arXiv{2006.06872} [hep-th]].

\bibitem{Page:1993wv}
D.~N.~Page,
``Information in black hole radiation,''
Phys. Rev. Lett. \textbf{71} (1993), 3743-3746
[\arXiv{hep-th/9306083} [hep-th]].

\bibitem{Page:2013dx}
D.~N.~Page,
``Time Dependence of Hawking Radiation Entropy,''
JCAP \textbf{09}, 028 (2013)
[\arXiv{1301.4995} [hep-th]].

\bibitem{Engelhardt:2014gca}
N.~Engelhardt and A.~C.~Wall,
``Quantum Extremal Surfaces: Holographic Entanglement Entropy beyond the Classical Regime,''
JHEP \textbf{01} (2015), 073
[\arXiv{1408.3203} [hep-th]].

\bibitem{Engelsoy:2016xyb}
  J.~Engelsöy, T.~G.~Mertens and H.~Verlinde,
  ``An investigation of AdS$_{2}$ backreaction and holography,''
  JHEP {\bf 1607} (2016) 139
  [\arXiv{1606.03438} [hep-th]].


\bibitem{Almheiri:2019psf}
  A.~Almheiri, N.~Engelhardt, D.~Marolf and H.~Maxfield,
  ``The entropy of bulk quantum fields and the entanglement wedge of an evaporating black hole,''
  JHEP {\bf 1912} (2019) 063
  [\arXiv{1905.08762} [hep-th]].

\bibitem{Penington:2019npb}
G.~Penington,
``Entanglement Wedge Reconstruction and the Information Paradox,''
JHEP \textbf{09} (2020), 002
[\arXiv{1905.08255} [hep-th]].

\bibitem{Almheiri:2019yqk}
  A.~Almheiri, R.~Mahajan and J.~Maldacena,
  ``Islands outside the horizon,''
  \arXiv{1910.11077} [hep-th].  
  
\bibitem{Ryu:2006bv}
  S.~Ryu and T.~Takayanagi,
  ``Holographic derivation of entanglement entropy from AdS/CFT,''
  Phys.\ Rev.\ Lett.\  {\bf 96} (2006) 181602
  [\arXiv{hep-th/0603001}].
 
 
\bibitem{Hubeny:2007xt}
  V.~E.~Hubeny, M.~Rangamani and T.~Takayanagi,
  ``A Covariant holographic entanglement entropy proposal,''
  JHEP {\bf 0707} (2007) 062
  [\arXiv{0705.0016} [hep-th]]. 

\bibitem{Faulkner:2013ana}
  T.~Faulkner, A.~Lewkowycz and J.~Maldacena,
  ``Quantum corrections to holographic entanglement entropy,''
  JHEP {\bf 1311} (2013) 074
  [\arXiv{1307.2892} [hep-th]].






\bibitem{Geng:2021iyq}
H.~Geng,
``Holographic BCFTs and Communicating Black Holes,''
[\arXiv{2104.07039} [hep-th]].

\bibitem{Bhattacharya:2021jrn}
A.~Bhattacharya, A.~Bhattacharyya, P.~Nandy and A.~K.~Patra,
``Islands and complexity of eternal black hole and radiation subsystems for a doubly holographic model,''
[\arXiv{2103.15852} [hep-th]].

\bibitem{Kawabata:2021hac}
K.~Kawabata, T.~Nishioka, Y.~Okuyama and K.~Watanabe,
``Probing Hawking radiation through capacity of entanglement,''
[\arXiv{2102.02425} [hep-th]].

\bibitem{Bousso:2021sji}
R.~Bousso and A.~Shahbazi-Moghaddam,
``Island Finder and Entropy Bound,''
[\arXiv{2101.11648} [hep-th]].

\bibitem{Wang:2021woy}
X.~Wang, R.~Li and J.~Wang,
``Quantifying islands and Page curves of Reissner-Nordstr\"om black holes for resolving information paradox,''
[\arXiv{2101.06867} [hep-th]].

\bibitem{Karananas:2020fwx}
G.~K.~Karananas, A.~Kehagias and J.~Taskas,
``Islands in Linear Dilaton Black Holes,''
[\arXiv{2101.00024} [hep-th]].

\bibitem{Hayden:2020vyo}
P.~Hayden and G.~Penington,
``Black hole microstates vs. the additivity conjectures,''
[\arXiv{2012.07861} [hep-th]].

\bibitem{Basak:2020aaa}
J.~Kumar Basak, D.~Basu, V.~Malvimat, H.~Parihar and G.~Sengupta,
``Islands for Entanglement Negativity,''
[\arXiv{2012.03983} [hep-th]].

\bibitem{Choudhury:2020hil}
S.~Choudhury, S.~Chowdhury, N.~Gupta, A.~Mishara, S.~P.~Selvam, S.~Panda, G.~D.~Pasquino, C.~Singha and A.~Swain,
``Circuit Complexity From Cosmological Islands,''
[\arXiv{2012.10234} [hep-th]].

\bibitem{Colin-Ellerin:2020mva}
S.~Colin-Ellerin, X.~Dong, D.~Marolf, M.~Rangamani and Z.~Wang,
``Real-time gravitational replicas: Formalism and a variational principle,''
[\arXiv{2012.00828} [hep-th]].

\bibitem{Goto:2020wnk}
K.~Goto, T.~Hartman and A.~Tajdini,
``Replica wormholes for an evaporating 2D black hole,''
[\arXiv{2011.09043} [hep-th]].

\bibitem{Matsuo:2020ypv}
Y.~Matsuo,
``Islands and stretched horizon,''
[\arXiv{2011.08814} [hep-th]].

\bibitem{Hernandez:2020nem}
J.~Hernandez, R.~C.~Myers and S.~M.~Ruan,
``Quantum extremal islands made easy. Part III. Complexity on the brane,''
JHEP \textbf{02} (2021), 173
[\arXiv{2010.16398} [hep-th]].

\bibitem{Bhattacharya:2020uun}
A.~Bhattacharya, A.~Chanda, S.~Maulik, C.~Northe and S.~Roy,
``Topological shadows and complexity of islands in multiboundary wormholes,''
JHEP \textbf{02}, 152 (2021)
[\arXiv{2010.04134} [hep-th]].

\bibitem{Ling:2020laa}
Y.~Ling, Y.~Liu and Z.~Y.~Xian,
``Island in Charged Black Holes,''
[\arXiv{2010.00037} [hep-th]].

\bibitem{Chen:2020hmv}
H.~Z.~Chen, R.~C.~Myers, D.~Neuenfeld, I.~A.~Reyes and J.~Sandor,
``Quantum Extremal Islands Made Easy, Part II: Black Holes on the Brane,''
JHEP \textbf{12} (2020), 025
[\arXiv{2010.00018} [hep-th]].

\bibitem{Johnson:2020mwi}
C.~V.~Johnson,
``Low Energy Thermodynamics of JT Gravity and Supergravity,''
[\arXiv{2008.13120} [hep-th]].

\bibitem{Chen:2020jvn}
H.~Z.~Chen, Z.~Fisher, J.~Hernandez, R.~C.~Myers and S.~M.~Ruan,
``Evaporating Black Holes Coupled to a Thermal Bath,''
JHEP \textbf{01} (2021), 065
[\arXiv{2007.11658} [hep-th]].

\bibitem{Chandrasekaran:2020qtn}
V.~Chandrasekaran, M.~Miyaji and P.~Rath,
``Including contributions from entanglement islands to the reflected entropy,''
Phys. Rev. D \textbf{102} (2020) no.8, 086009
[\arXiv{2006.10754} [hep-th]].

\bibitem{Li:2020ceg}
T.~Li, J.~Chu and Y.~Zhou,
``Reflected Entropy for an Evaporating Black Hole,''
JHEP \textbf{11} (2020), 155
[\arXiv{2006.10846} [hep-th]].

\bibitem{Chen:2020uac}
H.~Z.~Chen, R.~C.~Myers, D.~Neuenfeld, I.~A.~Reyes and J.~Sandor,
``Quantum Extremal Islands Made Easy, Part I: Entanglement on the Brane,''
JHEP \textbf{10} (2020), 166
[\arXiv{2006.04851} [hep-th]].

\bibitem{Alishahiha:2020qza}
M.~Alishahiha, A.~Faraji Astaneh and A.~Naseh,
``Island in the presence of higher derivative terms,''
JHEP \textbf{02}, 035 (2021)
[\arXiv{2005.08715} [hep-th]].

\bibitem{Hashimoto:2020cas}
K.~Hashimoto, N.~Iizuka and Y.~Matsuo,
``Islands in Schwarzschild black holes,''
JHEP \textbf{06} (2020), 085
[\arXiv{2004.05863} [hep-th]].

\bibitem{Giddings:2020yes}
S.~B.~Giddings and G.~J.~Turiaci,
``Wormhole calculus, replicas, and entropies,''
JHEP \textbf{09} (2020), 194
[\arXiv{2004.02900} [hep-th]].

\bibitem{Anegawa:2020ezn}
T.~Anegawa and N.~Iizuka,
``Notes on islands in asymptotically flat 2d dilaton black holes,''
JHEP \textbf{07}, 036 (2020)
[\arXiv{2004.01601} [hep-th]].

\bibitem{Gautason:2020tmk}
F.~F.~Gautason, L.~Schneiderbauer, W.~Sybesma and L.~Thorlacius,
``Page Curve for an Evaporating Black Hole,''
JHEP \textbf{05} (2020), 091
[\arXiv{2004.00598} [hep-th]].

\bibitem{Chen:2020wiq}
Y.~Chen, X.~L.~Qi and P.~Zhang,
``Replica wormhole and information retrieval in the SYK model coupled to Majorana chains,''
JHEP \textbf{06} (2020), 121
[\arXiv{2003.13147} [hep-th]].

\bibitem{Bhattacharya:2020ymw}
A.~Bhattacharya,
``Multipartite purification, multiboundary wormholes, and islands in $AdS_3/CFT_2$,''
Phys. Rev. D \textbf{102}, no.4, 046013 (2020)
[\arXiv{2003.11870} [hep-th]].


\bibitem{Chen:2019iro}
Y.~Chen,
``Pulling Out the Island with Modular Flow,''
JHEP \textbf{03} (2020), 033
[\arXiv{1912.02210} [hep-th]].




\bibitem{Almheiri:2019hni}
A.~Almheiri, R.~Mahajan, J.~Maldacena and Y.~Zhao,
``The Page curve of Hawking radiation from semiclassical geometry,''
JHEP \textbf{03} (2020), 149
[\arXiv{1908.10996} [hep-th]].

\bibitem{Marolf:2020xie}
D.~Marolf and H.~Maxfield,
``Transcending the ensemble: baby universes, spacetime wormholes, and the order and disorder of black hole information,''
JHEP \textbf{08} (2020), 044
[\arXiv{2002.08950} [hep-th]].

\bibitem{Marolf:2020rpm}
D.~Marolf and H.~Maxfield,
``Observations of Hawking radiation: the Page curve and baby universes,''
[\arXiv{2010.06602} [hep-th]].


\bibitem{Liu:2020jsv}
H.~Liu and S.~Vardhan,
``Entanglement entropies of equilibrated pure states in quantum many-body systems and gravity,''
P. R. X. Quantum. \textbf{2} (2021), 010344
[\arXiv{2008.01089} [hep-th]].

\bibitem{Pollack:2020gfa}
J.~Pollack, M.~Rozali, J.~Sully and D.~Wakeham,
``Eigenstate Thermalization and Disorder Averaging in Gravity,''
Phys. Rev. Lett. \textbf{125} (2020) no.2, 021601
[\arXiv{2002.02971} [hep-th]].

\bibitem{Sasieta:2021pzj}
M.~Sasieta,
``Ergodic Equilibration of R\'enyi Entropies and Replica Wormholes,''
[\arXiv{2103.09880} [hep-th]].

\bibitem{Krishnan:2021faa}
C.~Krishnan and V.~Mohan,
``Hints of Gravitational Ergodicity: Berry's Ensemble and the Universality of the Semi-Classical Page Curve,''
[\arXiv{2102.07703} [hep-th]].

\bibitem{Jackiw:1984je}
  R.~Jackiw,
  ``Lower Dimensional Gravity,''
  Nucl.\ Phys.\ B {\bf 252} (1985) 343.


\bibitem{Teitelboim:1983ux}
  C.~Teitelboim,
  ``Gravitation and Hamiltonian Structure in Two Space-Time Dimensions,''
  Phys.\ Lett.\  {\bf 126B} (1983) 41.

\bibitem{Almheiri:2019psy}
A.~Almheiri, R.~Mahajan and J.~E.~Santos,
``Entanglement islands in higher dimensions,''
SciPost Phys. \textbf{9}, no.1, 001 (2020)
[\arXiv{1911.09666} [hep-th]].


 
 


\bibitem{Hollowood:2020cou}
T.~J.~Hollowood and S.~P.~Kumar,
``Islands and Page Curves for Evaporating Black Holes in JT Gravity,''
JHEP \textbf{08} (2020), 094
[\arXiv{2004.14944} [hep-th]].

\bibitem{Hollowood:2020kvk}
T.~J.~Hollowood, S.~Prem Kumar and A.~Legramandi,
``Hawking radiation correlations of evaporating black holes in JT gravity,''
J. Phys. A \textbf{53} (2020) no.47, 475401
[\arXiv{2007.04877} [hep-th]].

 
\bibitem{Brown:2019rox}
A.~R.~Brown, H.~Gharibyan, G.~Penington and L.~Susskind,
``The Python\textquoteright{}s Lunch: geometric obstructions to decoding Hawking radiation,''
JHEP \textbf{08} (2020), 121
[\arXiv{1912.00228} [hep-th]].

\bibitem{Maldacena:2013xja}
J.~Maldacena and L.~Susskind,
``Cool horizons for entangled black holes,''
Fortsch. Phys. \textbf{61} (2013), 781-811
[\arXiv{1306.0533} [hep-th]].


\bibitem{Raju:2020smc}
S.~Raju,
``Lessons from the Information Paradox,''
[\arXiv{2012.05770} [hep-th]].

\bibitem{Laddha:2020kvp}
A.~Laddha, S.~G.~Prabhu, S.~Raju and P.~Shrivastava,
``The Holographic Nature of Null Infinity,''
SciPost Phys. \textbf{10}, 041 (2021)
[\arXiv{2002.02448} [hep-th]].

\bibitem{Geng:2020fxl}
H.~Geng, A.~Karch, C.~Perez-Pardavila, S.~Raju, L.~Randall, M.~Riojas and S.~Shashi,
``Information Transfer with a Gravitating Bath,''
[\arXiv{2012.04671} [hep-th]].


\bibitem{Geng:2020qvw}
H.~Geng and A.~Karch,
``Massive islands,''
JHEP \textbf{09}, 121 (2020)
\arXiv{2006.02438} [hep-th]].

\bibitem{Geng:2021hlu}
H.~Geng, A.~Karch, C.~Perez-Pardavila, S.~Raju, L.~Randall, M.~Riojas and S.~Shashi,
``Inconsistency of Islands in Theories with Long-Range Gravity,''
[\arXiv{2107.03390} [hep-th]].

\bibitem{Ghosh:2021axl}
K.~Ghosh and C.~Krishnan,
``Dirichlet baths and the not-so-fine-grained Page curve,''
JHEP \textbf{08}, 119 (2021)
[\arXiv{2103.17253} [hep-th]].

\bibitem{BD}
N.~D.~Birrel and P.~C.~W.~Davies, ``Quantum fields in curved space," Cambridge University Press, 1982.


\bibitem{Jacobson:2003vx}
T.~Jacobson,
``Introduction to quantum fields in curved space-time and the Hawking effect,''
[\arXiv{gr-qc/0308048}].

\bibitem{Araki:1970ba}
H.~Araki and E.~Lieb,
``Entropy inequalities,''
Commun. Math. Phys. \textbf{18} (1970), 160-170


\bibitem{CW}
M. Christandl and A. Winter, 
"Squashed entanglement: An additive entanglement measure,"
J. Math. Phys. \textbf{45.3} (2004) 829
[\arXiv{quant-ph/0308088}].


\bibitem{Hayden:2007cs}
  P.~Hayden and J.~Preskill,
  ``Black holes as mirrors: Quantum information in random subsystems,''
  JHEP {\bf 0709} (2007) 120
  [\arXiv{0708.4025} [hep-th]].

\bibitem{Harlow:2013tf}
D.~Harlow and P.~Hayden,
``Quantum Computation vs. Firewalls,''
JHEP \textbf{06} (2013), 085
[\arXiv{1301.4504} [hep-th]]. 


\bibitem{Grey}
T.~J.~Hollowood and N.~Talwar, ``Greybody Factors, Islands and Correlations in Hawking  Radiation,''
{\sl to appear\/}.
  
  
\bibitem{Calabrese:2004eu} 
  P.~Calabrese and J.~L.~Cardy,
  ``Entanglement entropy and quantum field theory,''
  J.\ Stat.\ Mech.\  {\bf 0406}, P06002 (2004)
  [\arXiv{hep-th/0405152}].

\bibitem{Akers:2019lzs}
C.~Akers, N.~Engelhardt, G.~Penington and M.~Usatyuk,
``Quantum Maximin Surfaces,''
JHEP \textbf{08}, 140 (2020)
[\arXiv{1912.02799} [hep-th]].

}
\end{thebibliography}
\end{document}